\documentclass{article}

\usepackage{times}
\usepackage{graphicx} 
\usepackage{subfigure} 

\usepackage{natbib}

\usepackage{amsmath}
\usepackage{amsfonts}
\usepackage{amssymb}
\usepackage{amsthm}
\usepackage{mathtools}
\usepackage{bm}

\usepackage{enumitem}
\newcommand{\subscript}[2]{$#1 _ #2$}

\usepackage{algorithm}
\usepackage{algorithmic}

\newtheorem{theorem}{Theorem}

\newtheorem{corollary}{Corollary}
\newtheorem{proposition}{Proposition}

\DeclareMathOperator*{\argmin}{arg\,min}
\DeclareMathOperator*{\argmax}{arg\,max}
\DeclarePairedDelimiter\ceil{\lceil}{\rceil}

\usepackage{hyperref}

\usepackage[accepted]{icml2018}

\icmltitlerunning{Bayesian Quadrature for Multiple Related Integrals}

\begin{document} 

\twocolumn[
\icmltitle{Bayesian Quadrature for Multiple Related Integrals}



\icmlsetsymbol{equal}{*}

\begin{icmlauthorlist}
\icmlauthor{Xiaoyue Xi}{imperial,equal}
\icmlauthor{Fran\c{c}ois-Xavier Briol}{imperial,warwick,ATI,equal}
\icmlauthor{Mark Girolami}{imperial,warwick,ATI}
\end{icmlauthorlist}

\icmlaffiliation{imperial}{Department of Mathematics, Imperial College London}
\icmlaffiliation{warwick}{Department of Statistics, University of Warwick}
\icmlaffiliation{ATI}{The Alan Turing Institute for Data Science and AI}

\icmlcorrespondingauthor{Fran\c{c}ois-Xavier Briol}{f-x.briol@warwick.ac.uk}


\vskip 0.3in
]



\printAffiliationsAndNotice{\icmlEqualContribution} 

\begin{abstract} 
Bayesian probabilistic numerical methods are a set of tools providing posterior distributions on the output of numerical methods. The use of these methods is usually motivated by the fact that they can represent our uncertainty due to incomplete/finite information about the continuous mathematical problem being approximated. In this paper, we demonstrate that this paradigm can provide additional advantages, such as the possibility of transferring information between several numerical methods. This allows users to represent uncertainty in a more faithful manner and, as a by-product, provide increased numerical efficiency. We propose the first such numerical method by extending the well-known Bayesian quadrature algorithm to the case where we are interested in computing the integral of several related functions. We then prove convergence rates for the method in the well-specified and misspecified cases, and demonstrate its efficiency in the context of multi-fidelity models for complex engineering systems and a problem of global illumination in computer graphics.
\end{abstract} 

\section{Introduction}
\label{sec:introduction}

Probabilistic numerics \citep{Hennig2015} proposes approaching problems of numerical analysis from the point of view of statistics. In particular, Bayesian probabilistic numerical methods approach this problem from a Bayesian point of view, and can provide posterior distributions on the solutions of numerical problems (e.g. in the case of this paper, the solution of some integral). These posterior distributions represent our epistemic uncertainty about these quantities of interest. In the case of quadrature rules, the uncertainty is due to the fact that we only have a finite number of function evaluations and therefore are uncertaint about the value of the integral. The notion of \emph{Bayesian probabilistic numerical method} was independently introduced by several authors \citep{Larkin1972,Kadane1985,Diaconis1988,OHagan1992}, but only recently formalised by \cite{Cockayne2017}. 

Apart from the uncertainty quantification property described above, these methods have several other advantages over ``classical" (i.e. non-Bayesian) numerical methods (although some of the classical and Bayesian methods coincide \citep{Diaconis1988}). First of all, they allow the user to formulate all of its prior knowledge in the form of a prior, making all of the assumptions of the numerical scheme explicit. Second of all, they can allow for coherent propagation of numerical uncertainties through chains of computation; see \citep{Cockayne2017,Oates2017hydrocyclones}.

However, one property which has not been studied so far is the possibility of jointly inferring several quantities of interest. In this paper, we study the problem of numerically integrating a sequence of functions $f_1,\ldots,f_D$ (which are correlated to one another) with respect to some probability measure $\Pi$, and hence propose to build a model for joint inference of $\int f_1 \mathrm{d}\Pi,\ldots,\int f_D \mathrm{d}\Pi$. Such a joint model allows for better finite-sample performance, and can also lead to more refined posterior distributions on each of the individual integrals.

To tackle this problem, we extend the well-known Bayesian quadrature \citep{OHagan1991} algorithm and study the performance of the proposed methodology from a theoretical and experimental point of view. In particular, we provide asymptotic convergence results for the marginal posterior variance on each of the integrals, both in the case of a well specified and misspecified prior. We also demonstrate the performance of our algorithm on some toy problems from the engineering literature on multi-fidelity models, and on a challenging problem from the field of computer graphics.

\section{Methodology} 

\paragraph{Bayesian Quadrature}

Let $(\mathcal{X},\mathcal{B},\Pi)$ be a probability space and consider some function $f:\mathcal{X}\rightarrow \mathbb{R}$ where $\mathcal{X}\subseteq \mathbb{R}^p, p \in \mathbb{N}_+$. The classical problem of numerical integration is concerned with approximating the integral:
\begin{eqnarray*}
\Pi[f] & := & \int_{\mathcal{X}} f(\bm{x}) \Pi(\mathrm{d}\bm{x}),
\end{eqnarray*}
where we assume $\int_{\mathcal{X}} f^2(\bm{x}) \Pi(\mathrm{d}\bm{x}) < \infty$. Under fairly general conditions on $f$, one can show that an optimal algorithm (in terms of worst-case integration error in some function space) takes the form of a quadrature (or cubature) rule $\hat{\Pi}[f] = \sum_{i=1}^N w_i f(\bm{x}_i)$ for some weights $\{w_i\}_{i=1}^N \in \mathbb{R}$ and samples $\{\bm{x}_i\}_{i=1}^N \in \mathcal{X}$ (see \cite{Bakhvalov1971}). These are also sometimes denoted in vectorised form as $\Pi[f]=\bm{w}^\top f(\bm{X})$ where $\bm{w}=(w_1,\ldots,w_N)^\top$, $\bm{X}=(\bm{x}_1,\ldots,\bm{x}_N)^\top$ and $f(\bm{X})=(f(\bm{x}_1),\ldots,f(\bm{x}_N))^\top$. The notation $\hat{\Pi}[f]$ is motivated by the fact that we can see this object as an exact integral with respect to a discrete measure $\hat{\Pi}=\sum_{i=1}^N w_i \delta_{\bm{x}_i}$, where $\delta_{\bm{x}_i}$ denotes the Dirac delta measure taking value $1$ at $\bm{x}_i$ and $0$ otherwise. Many popular numerical integration methods take this form, including Newton--Cotes rules, Gaussian quadrature, Monte Carlo methods and sparse grids.

Let $(\Omega,\mathcal{F},\mathbb{P})$ be another probability space. \textit{Bayesian quadrature} (BQ), introduced by \cite{OHagan1991}, proposes to approach the problem of numerical integration by first formulating a prior stochastic model $g:\mathcal{X}\times \Omega \rightarrow \mathbb{R}$  for the integrand $f$ (where $\forall \omega \in \Omega$, $g(\cdot,\omega)$ represents a realisation of $g$). This prior model is then conditioned on the vector of observations $f(\bm{X})$ to obtain a posterior model for $f$, which is then pushed forward through the integral operator to give a posterior on $\Pi[f]$.

A popular choice of prior is a Gaussian Process (GP) $\mathcal{GP}(m,k)$ with $m:\mathcal{X}\rightarrow \mathbb{R}$ denoting the mean function (i.e. $m(\bm{x}) = \mathbb{E}_{\omega}[g(\bm{x},\omega)]$), and $c:\mathcal{X}\times \mathcal{X}\rightarrow \mathbb{R}$ denoting the covariance function/kernel (i.e. $c(\bm{x},\bm{x}') = \mathbb{E}_{\omega}[(g(\bm{x},\omega)-m(\bm{x}))(g(\bm{x}',\omega)-m(\bm{x}'))]$). Let us assume that $m=0$ (this can be done without loss of generality since the domain can be re-parametrized to be centred at $0$). After conditioning on $X$, we have a new Gaussian process $g_N$ with mean and covariance:
\begin{eqnarray*}
m_N(\bm{x}) & = & c(\bm{x},\bm{X})c(\bm{X},\bm{X})^{-1}f(\bm{X}), \\
c_N(\bm{x},\bm{x}') & = & c(\bm{x},\bm{x}') - c(\bm{x},\bm{X})c(\bm{X},\bm{X})^{-1}c(\bm{X},\bm{x}'),
\end{eqnarray*}
for all $\bm{x},\bm{x}' \in \mathcal{X}$. Here, $c(\bm{X},\bm{X})$ is the Gram matrix with entries $(c(\bm{X},\bm{X}))_{ij} = c(\bm{x}_i,\bm{x}_j)$ and $c(\bm{x},\bm{X})=(c(\bm{x},\bm{x}_1),\ldots,c(\bm{x},\bm{x}_N))$ whilst $c(\bm{X},\bm{x}) = c(\bm{x},\bm{X})^\top$ . The push-forward of this posterior through the integral operator is a Gaussian distribution with mean and variance: 
\begin{eqnarray*}
\mathbb{E}\left[\Pi[g_N]\right] & = &  \Pi[c(\cdot,\bm{X})]c(\bm{X},\bm{X})^{-1} f(\bm{X}),\\
\mathbb{V}\left[\Pi[g_N]\right] & = & \Pi\bar{\Pi} \left[c\right] - \Pi[c(\cdot,\bm{X})]c(\bm{X},\bm{X})^{-1}\bar{\Pi}[c(\bm{X},\cdot)],
\end{eqnarray*}
where $\Pi[c(\cdot,\bm{X})] = (\Pi[c(\cdot,\bm{x}_1)],\ldots,\Pi[c(\cdot,\bm{x}_N)])$. These expression can be obtained in closed-form if the \textit{kernel mean} $\Pi[c(\cdot,\bm{x})]=\int_{\mathcal{X}} c(\bm{x}',\bm{x}) \Pi(\mathrm{d}\bm{x}')$ (also called the representer of integration) and \textit{initial error} $\Pi\bar{\Pi}[c] = \int_{\mathcal{X}\times \mathcal{X}} c(\bm{x},\bm{x}') \Pi(\mathrm{d}\bm{x}) \Pi(\mathrm{d}\bm{x}')$ can be obtained in closed form (here $\bar{\Pi}$ indicates that the integral is taken with respect to the second argument). 

The choice of covariance function $c$ can be used to encode prior beliefs about the function $f$, such as smoothness or periodicity, and is very important to obtain good performance in practice. A popular example is the family of Mat\'ern kernels 
\begin{equation}\label{eq:matern_kernels}
\begin{split}
c_{\alpha}(\bm{x},\bm{x}') \; & = \; \lambda^2 \frac{2^{1-\alpha}}{\Gamma(\alpha)} \left(\sqrt{2 \alpha} \frac{\|\bm{x}-\bm{x}'\|_{2}^{2}}{\sigma^2}\right)^{\alpha}  \\
&   \qquad \qquad \qquad \times J_{\alpha} \left(\sqrt{2 \alpha}  \frac{\|\bm{x}-\bm{x}'\|_{2}^{2}}{\sigma^2}\right),
\end{split}
\end{equation}
for $\sigma,\lambda>0$ where $J_{\alpha}$ is the Bessel function of the second kind and $\alpha>0$ gives the smoothness of the kernel. On $\mathcal{X}=\mathbb{R}^p$, this will give an RKHS norm-equivalent to the Sobolev space $W_2^{\alpha}(\mathbb{R}^d)$\footnote{We say that two norms $\|\cdot\|_{1}$ and $\|\cdot\|_{2}$ on a vector space are \emph{norm-equivalent} if and only if $\exists C_1,C_2>0$ such that $C_1 \|\cdot\|_{2} \leq \|\cdot\|_{1} \leq C_2 \|\cdot\|_{2}$.}. Examples of infinitely smooth kernels include the squared-exponential kernel $c(\bm{x},\bm{x}') = \exp(-\|\bm{x}-\bm{x}'\|_2^2/\sigma^2)$ where $\sigma>0$, the multi-quadric kernel $c(\bm{x},\bm{x}') = (-1)^{\ceil{\beta}}(\sigma^2+\|\bm{x}-\bm{x}'\|_2^2)^{\beta}$ for $\beta,\sigma>0, \beta \not\in \mathbb{N}$ and the inverse multi-quadric kernel $c(\bm{x},\bm{x}') = (\sigma^2+ \|\bm{x}-\bm{x}'\|_2^2)^{-\beta}$ for $\beta,\sigma>0$.

In practice, numerical inversion can be challenging since the Gram matrix tends to be nearly singular, and so one may wish to regularise the matrix using a Tikhonov penalty. The inverses above can also potentially render the computation of the BQ estimator computationally expensive (up to $\mathcal{O}(N^3)$ cost in the most general settings), although this can be alleviated in specific cases \cite{Karvonen2017symm}. Even if this is not the case, the additional cost can be worthwhile regardless since the method has been shown to attain fast convergence rates \cite{Briol2015,Briol2015a,Kanagawa2016,Kanagawa2017,Bach2015} when the target integrand and the kernel used are smooth. 

Recent research directions in BQ include efficient sampling algorithms (for the point set $\bm{X}$) to improve the performance of the method \cite{Rasmussen2003,Huszar2012,Gunter2014,Briol2015,Karvonen2017,Briol2017}, asymptotic convergence results \cite{Briol2015,Briol2015a,Kanagawa2016,Bach2015} and equivalence of BQ with known quadrature rules for certain choices of point sets and kernels \cite{Sarkka2015,Karvonen2017}. Furthermore, there has also been a wide range of new applications, including to other numerical methods in optimization, linear algebra and functional approximation \cite{Kersting2016,Fitzsimons2017}, inference in complex computer models \cite{Oates2016}, and problems in econometrics \cite{Oettershagen2017} and computer graphics \cite{Brouillat2009,Marques2013,Briol2015a}.

Although other stochastic processes could of course be used \citep{Cockayne2017}, GPs are popular due to their conjugacy properties, and the terminology Bayesian quadrature usually refers to this case. Note that other names for BQ with GP priors include Gaussian-process quadrature \citep{Sarkka2015} or kernel quadrature \citep{Bach2015,Briol2017,Kanagawa2017}. In fact, a well-known alternative view of the posterior mean provided by BQ is that of an optimally-weighted quadrature rule in a reproducing kernel Hilbert spaces (RKHS) in the classical worst-case setting \cite{Ritter2000}. Let $\mathcal{H}_k$ be an RKHS with inner product and norm denoted $\langle\cdot,\cdot\rangle_k$ and $\|\cdot\|_k$ respectively; i.e. a Hilbert space with an associated symmetric and positive definite reproducing kernel $k:\mathcal{X}\times \mathcal{X} \rightarrow \mathbb{R}$ such that $f(\bm{x}) = \langle f, k(\cdot,\bm{x})\rangle_k$ (see \cite{Berlinet2004} for a detailed study). Suppose that our integrand $f\in \mathcal{H}_k$ and that $\int_{\mathcal{X}} k(\bm{x},\bm{x}) \Pi(\mathrm{d}\bm{x}) < \infty$. In that case, using the Cauchy--Schwarz inequality, the integration error can be decomposed as:
\begin{eqnarray*}
\left|\Pi[f] - \hat{\Pi}[f]\right| & \leq & \left\|f\right\|_{k} \left\| \Pi\left[k(\cdot,\bm{x})\right] - \hat{\Pi}\left[k(\cdot,\bm{x})\right] \right\|_{k}.
\end{eqnarray*}
The corresponding worst-case error over the unit ball of the space $\mathcal{H}_k$ is given by:
\begin{multline*}
e \left(\mathcal{H}_k,\hat{\Pi},\bm{X}\right) 
  \; = \; \sup_{\|f\|_k \leq 1} \left|\Pi[f] - \hat{\Pi}[f]\right| \\
 =  \; \left\| \Pi\left[k(\cdot,\bm{x})\right] - \hat{\Pi}\left[k(\cdot,\bm{x})\right] \right\|_{k} \hspace{23mm} \\
 =  \; \Big(\bm{w}^\top k(\bm{X},\bm{X}) \bm{w} - 2\Pi[k(\cdot,\bm{X})]^\top \bm{w} + \Pi \bar{\Pi}[k]\Big)^{\frac{1}{2}}.
\end{multline*}
This final expression can be minimised in closed form over $\bm{w} \in \mathbb{R}^{N}$ to show that the optimal quadrature rule has weights $\bm{w} = \Pi[k(\cdot,\bm{X})] k(\bm{X},\bm{X})^{-1}$. This corresponds exactly to the weights for the BQ posterior mean if we take our prior on $f$ to be a $\mathcal{GP}(0,k)$, whilst the worst-case error can be shown to correspond to the posterior variance squared. The BQ estimator with prior $\mathcal{GP}(0,c)$ is therefore optimal in the classical worst-case sense for the RKHS $\mathcal{H}_c$. 

\paragraph{Multi-output Bayesian Quadrature}\label{sec:multioutput_BQ}

We now extend the set-up of our problem. Suppose we have a sequence of probability spaces $(X_d,\mathcal{B}_d,\Pi_d)$ and functions $f_d:\mathcal{X}_d \rightarrow \mathbb{R}$ for which we are interested in numerically computing integrals of the form $\Pi_d[f_d]$ for $d=1,\ldots,D$.  In many applications where we are faced with this type of problem, we also have prior knowledge about correlations between the individual $f_d$. However, this information is often ignored and the integrals are approximated individually. This is not a principled approach from a Bayesian point of view since it means we are not conditioning on all available information. In this section, we extend the BQ algorithm to solve this problem by building a joint model of $f_1,\ldots,f_D$ in order to obtain a joint posterior on the integrals $\Pi_1[f_1],\ldots,\Pi_D[f_D]$.

For notational convenience, we will restrict ourselves to the case where all of the input domains are identical and denoted $\mathcal{X}$, all of the probability measures are identical and denoted $\Pi$, and the input sets $\bm{X} = \{\bm{X}_d\}_{d=1}^D$ consist of $N$ points $\bm{X}_d = (\bm{x}_{d,1},\ldots,\bm{x}_{d,N})$ per output function $f_d$ (note the setup can be made more general if necessary). We re-frame the integration problem as that of integrating some vector-valued function $\bm{f}:\mathcal{X} \rightarrow \mathbb{R}^D$ such that $\bm{f}(\bm{x}) = (f_1(\bm{x}),\ldots,f_D(\bm{x}))^\top$; i.e. we want to estimate $\Pi[\bm{f}] = (\Pi[f_1],\ldots,\Pi[f_D])^\top$. In this multiple-integral setting, we can have generalised quadrature rules of the form:
\begin{eqnarray*}
\hat{\Pi}[f_d] & = & \sum_{d'=1}^D \sum_{i=1}^N (\bm{W}_{i})_{dd'} f_{d'}(\bm{x}_{d',i})
\end{eqnarray*}
where $\bm{W}_{i} \in \mathbb{R}^{D \times D}$ are weight matrices and $(\bm{W}_{i})_{dd'}$ gives the influence of the value of $f_{d'}$ at $\bm{x}_{d',i}$ on the estimate of $\Pi[f_d]$. The quadrature rule for $\bm{f}$ can be re-written in compact form as $\hat{\Pi}[\bm{f}] = \bm{W}^\top \bm{f}(\bm{X})$ for some weight matrix $\bm{W} \in \mathbb{R}^{ND \times D}$ (a concatenation of $\{\bm{W}_i\}_{i=1}^N$) and function-evaluations vector $\bm{f}(\bm{X}) = (f_1(\bm{x}_{1,1}),\ldots,f_1(\bm{x}_{1,N}),\ldots,f_D(\bm{x}_{D,1}),\ldots,f_D(\bm{x}_{D,N}))^\top$. 

These generalised quadrature rules encompass popular Monte Carlo methods such as control variates or functionals \cite{Glasserman2004,Oates2017}, multilevel Monte Carlo \cite{Giles2015} and multi-fidelity Monte Carlo \cite{Peherstorfer2016MC}. However, it is important to point out that these methods can only deal with very specific relations between integrands, usually requiring $(\int_{\mathcal{X}} (f_d(\bm{x})-f_{d'}(\bm{x}))^2 \Pi(\mathrm{d}\bm{x}))^{\frac{1}{2}}$ to be small for all pairs of integrands $f_d,f_{d'}$. Our method will be able to make use of much more complex relations. 

We propose to approach this problem using an extended version of BQ, where we impose a prior $\bm{g}:\mathcal{X}\times \Omega \rightarrow \mathbb{R}^D$ which is a $\mathcal{GP}(\bm{0},\bm{C})$ on the extended space (this is often called a multi-output GP or co-kriging model \cite{Alvarez2011}) where now $\bm{C}$ is matrix-valued and $(C(\bm{x},\bm{x}'))_{dd'} = \mathbb{E}_{\omega \sim \mathbb{P}}[g_d(\bm{x},\omega)g_{d'}(\bm{x}',\omega)]$. In this case, after conditioning on $\bm{X}$, we have a GP $\bm{g}_N$ with vector-valued mean $\bm{m}_N:\mathcal{X}\rightarrow \mathbb{R}^D$ and matrix-valued covariance $\bm{C}_N: \mathcal{X} \times \mathcal{X}\rightarrow \mathbb{R}^{D \times D}$:
\begin{align*}
\bm{m}_N(\bm{x}) & =  \bm{C}(\bm{x},\bm{X})\bm{C}(\bm{X},\bm{X})^{-1} \bm{f}(\bm{X}), \\
\bm{C}_N(\bm{x},\bm{x}')  & =  \bm{C}(\bm{x},\bm{x}')  - \bm{C}(\bm{x},\bm{X})\bm{C}(\bm{X},\bm{X})^{-1}\bm{C}(\bm{X},\bm{x}'),
\end{align*}
for $\bm{C}(\bm{x},\bm{X}) = (C(\bm{x},\bm{x}_1),\ldots,C(\bm{x},\bm{x}_N)) \in \mathbb{R}^{D \times ND}$ and Gram matrix $\bm{C}(\bm{X},\bm{X}) \in \mathbb{R}^{ND \times ND}$ is:

\begin{equation*}
\bm{C}(\bm{X},\bm{X}) = 
\begin{bmatrix}
    \left(\bm{C}(\bm{X}_1,\bm{X}_1)\right)_{1,1}  & \dots  & \left(\bm{C}(\bm{X}_1,\bm{X}_D)\right)_{1,D} \\
 \left(\bm{C}(\bm{X}_2,\bm{X}_1)\right)_{2,1} & \vdots & \left(\bm{C}(\bm{X}_2,\bm{X}_D)\right)_{2,D}  \\
    \vdots   & \vdots & \vdots \\
    \left(\bm{C}(\bm{X}_D,\bm{X}_1)\right)_{D,1}  & \dots  & \left(\bm{C}(\bm{X}_D,\bm{X}_D)\right)_{D,D}
\end{bmatrix},
\end{equation*}
where $C(\bm{X}_d,\bm{X}_{d'})_{d,d'}$ is an $N\times N$ matrix. The posterior on the value of the integral vector $\Pi[\bm{f}]$ can also be obtained whenever the kernel mean $\Pi[\bm{C}(\cdot,\bm{x})]$ and initial error $\Pi\bar{\Pi} \left[\bm{C}\right]$ are available in closed form, which is potentially a restrictive condition. The authors of \cite{Briol2015a} give a table of closed-form expressions of these quantities for popular kernels in the uni-output case, and we envision the same type of table being necessary for future extensions of multi-output BQ. Alternatively, \cite{Oates2017,Oates2016CF2} proposed a kernel which is tailored to the target probability measure $\Pi$ and which could also be extended to the multi-output case.
\begin{proposition}\label{prop:multioutput_BQ}
Consider multi-output Bayesian Quadrature with a $\mathcal{GP}(\bm{0},\bm{C})$ prior on $\bm{f}=(f_1,\ldots,f_D)^\top$. The posterior distribution on $\Pi[\bm{f}]$ is a $D$-dimensional Gaussian distribution with mean and covariance matrix:
\begin{eqnarray*}
\mathbb{E} \left[\Pi[\bm{g}_N]\right] & = & \Pi[\bm{C}(\cdot,\bm{X})]\bm{C}(\bm{X},\bm{X})^{-1} \bm{f}(\bm{X}),\\
\mathbb{V} \left[\Pi[\bm{g}_N]\right] & = & \Pi\bar{\Pi} \left[\bm{C}\right]  - \Pi[\bm{C}(\cdot,\bm{X})]\bm{C}(\bm{X},\bm{X})^{-1}\bar{\Pi}[\bm{C}(\bm{X},\cdot)].
\end{eqnarray*}
\end{proposition}
All proofs can be found in Appendix \ref{sec:appendix_proofs}. In this case, we clearly end up with a generalised quadrature rule with weight matrix:
$\bm{W}^{\text{BQ}} = (\Pi \left[\bm{C}(\cdot,\bm{X})\right]\bm{C}(\bm{X},\bm{X})^{-1})^{\top} \in \mathbb{R}^{ND \times D}$.
In general, the computational cost for computing the posterior mean and variance is now of order $\mathcal{O}(N^3D^3)$. However, several choices of kernels can reduce this cost significantly, and it is also possible to obtain sparse GP approximations; see e.g. \cite{Alvarez2011convolved}.

The choice of kernel $\bm{C}$ is of course once again of great importance since it encodes prior knowledge about each of the integrand and their correlation structure and should be made based on the application considered. We also remark that matrix valued kernels $\bm{C}$ can be described in term of some scalar-valued kernel $r$ on the extended space $\mathcal{X}\times\{1,\ldots,D\}$ as $(\bm{C}(\bm{x},\bm{x}'))_{dd'} = r((\bm{x},d),(\bm{x}',d'))$. We now present two choices of covariance functions which are popular in the literature and will be used in this paper:
\begin{itemize}
\item The \textit{separable} kernel is of the form 
\begin{eqnarray*}
\bm{C}(\bm{x},\bm{x}') & = & \bm{B} c(\bm{x},\bm{x}'),
\end{eqnarray*}
where $\bm{B} \in \mathbb{R}^{D \times D}$ is symmetric and positive definite, and $\bm{c}:\mathcal{X} \times \mathcal{X} \rightarrow \mathbb{R}$ is a scalar-valued reproducing kernel. This treats the kernel as the product of two scalar-valued reproducing kernels, one defined on $\mathcal{X}$ and the other on $\{1,\ldots,D\}$. A particular case of interest is the \textit{linear model of coregionalization} (LMC) where the matrix is of the form $(\bm{B})_{dd'} = \sum_{i=1}^{R} a^i_{d} a^i_{d'}$ for some $a^i_{d} \in \mathbb{R}$. This type of kernel can lead to a lower computational cost of order $\mathcal{O}(N^3 + D^3)$ when evaluating all $f_d$ on the same input set and using tensor product formulations (see Appendix \ref{sec:appendix_implementation}).

\item The \textit{process convolution} kernel \citep{VerHoef1998,Higdon2002,Alvarez2011} models the individual functions $f_1,\ldots,f_D$ as blurred transformations of $R \in \mathbb{N}_+$ underlying functions. It is given by:
\begin{eqnarray*}
(\bm{C}(\bm{x},\bm{x}'))_{d,d'} & = &  c_{d,d'}(\bm{x},\bm{x}') + c_{w_d}(\bm{x},\bm{x}') \delta_{d,d'}, 
\end{eqnarray*}
where $\delta_{dd'} = 1$ if $d=d'$ and $0$ else. Here there are two parts of the kernel, first $c_{d,d'}:\mathcal{X}\times \mathcal{X}\rightarrow \mathbb{R}$ defined as:
\begin{eqnarray*}
c_{d,d'}(\bm{x},\bm{x}')  & = & \sum_{i=1}^{R} \int_{\mathcal{X}} G^i_{d}(\bm{x}-\bm{z}) \times \\ & & \; \int_{\mathcal{X}} G^i_{d'}(\bm{x}'-\bm{z}') c_i(\bm{z},\bm{z}') \mathrm{d}\bm{z}' \mathrm{d}\bm{z},
\end{eqnarray*} 
and $c_{w_d}:\mathcal{X}\times \mathcal{X}\rightarrow \mathbb{R}$ representing covariance inherent to the $d^\text{th}$ function and $G_{d}^i:\mathcal{X}\rightarrow \mathbb{R}$ is a blurring kernel\footnote{Note that the term ``blurring kernel" does not mean the function is a reproducing kernel.} which is a continuous function either having compact support or being square integrable. Notice that taking $G^i_{d}(\bm{x}-\bm{z}) = a^i_{d} \delta(\bm{x}-\bm{z})$ (where $\delta(\cdot)$ represents a Dirac function) gives back the LMC case.
\end{itemize}
Note that it is also common to combine kernels, by summing them (i.e. $\bm{C}(\bm{x},\bm{x}') = \sum_{q=1}^Q \bm{C}_q(\bm{x},\bm{x}')$) in order to obtain more flexible models.  The kernel means and initial error, as well as other details for implementation are provided in Appendix \ref{sec:appendix_implementation}.

\section{Theoretical results}
\label{sec:MOBQ_theory}

In this section, we begin by exploring properties of multi-output BQ with $\mathcal{GP}(\bm{0},\bm{C})$ prior as an optimally-weighted quadrature algorithm in vector-valued RKHS $\mathcal{H}_{\bm{C}}$.

Let $\mathcal{H}_{\bm{K}}$ be a vector-valued RKHS with norm and inner product denoted $\|\cdot\|_{\bm{K}}$ and $\langle \cdot, \cdot \rangle_{\bm{K}}$ respectively. These spaces were extensively studied in \cite{Pedrick1957,Micchelli2008,Carmeli2006,Carmeli2008,DeVito2013}, and generalise the notion of RKHS to vector-valued functions. In the multi-output case, there is also a one-to-one correspondance between the RKHS $\mathcal{H}_{\bm{K}}$ and the kernel $\bm{K}$. Theorem 3.1 in \cite{Micchelli2008} shows that the minimizer of the variational problem:
\begin{eqnarray*}
\min_{\bm{h} \in \mathcal{H}_{\bm{K}}} \left\{ \|\bm{h}\|_{\bm{K}}^2 \; : \; \bm{h}:\mathcal{X}\rightarrow \mathbb{R}^D, \bm{h}(\bm{x}_i) = \bm{f}(\bm{x}_i) \; \forall \bm{x}_i \in \bm{X} \right\} 
\end{eqnarray*}
takes the form of the multi-output posterior GP mean $\bm{m}_N$ obtained after conditioning a $\mathcal{GP}(\bm{0},\bm{K})$ on some data set $\bm{X}$. We can therefore extend a well-known result from the uni-output case to show that $\hat{\Pi}_{\text{BQ}}[f_d]$ is an optimally weighted quadrature rule for all $f_d$ in terms of their worst-case integration error, denoted:
\begin{eqnarray}\label{eq:MOBQ_WCE}
e(\mathcal{H}_{\bm{C}},\hat{\Pi},\bm{X},d) 
& := & 
\sup_{\|\bm{f}\|_{\bm{C}}\leq 1} \left| \Pi[f_d] - \hat{\Pi}[f_d]\right|.
\end{eqnarray}  
\begin{proposition}[Optimally weighted quadrature rule in $\mathcal{H}_{\bm{C}}$]\label{prop:optimal_weights}
For a fixed point set $\bm{X}$, denote by $\hat{\Pi}[\bm{f}] = \bm{W}^\top \bm{f}(\bm{X})$ any quadrature rule for the vector-valued function $\bm{f}=(f_1,\ldots,f_D)$ and by $\hat{\Pi}_{\text{BQ}}[\bm{f}]=\bm{W}^\top_{\text{BQ}}\bm{f}(\bm{X})$ the BQ rule with $\mathcal{GP}(\bm{0},\bm{C})$ prior. Then, $\forall d=1,\ldots,D$:
\begin{eqnarray*} 
\bm{W}_{\text{BQ}} & = & \argmin_{\bm{W} \in \mathbb{R}^{ND \times D}} e(\mathcal{H}_{\bm{C}},\hat{\Pi},\bm{X},d).
\end{eqnarray*}
\end{proposition}
In specific cases, it is also possible to characterise the rate of convergence of the worst-case error for each element $f_d$. This is for example the case with the separable kernel introduced in Sec. \ref{sec:multioutput_BQ}, as will be demonstrated in the Theorem \ref{theorem:convergence_separable} below. First, we introduce some technical definitions which will be required for the statement of the theorem. 

We say that a domain $\mathcal{X} \subset \mathbb{R}^p$ satisfies an \textit{interior cone condition} if there exists an angle $\theta \in (0,\frac{\pi}{2})$ and a radius $r>0$ such that $\forall \bm{x} \in \mathcal{X}$, a unit vector $\bm{\xi}(\bm{x})$ exists such that the cone $\{\bm{x}+\lambda \bm{y}: \bm{y} \in \mathbb{R}^p, \|\bm{y}\|_2 =1, \bm{y}^\top \bm{\xi}(\bm{x}) \geq \cos \theta, \lambda \in [0,r]\}$ is a subset of $\mathcal{X}$. 

For a point set $\bm{X}$, we call $h_{\bm{X},\mathcal{X}} := \sup_{\bm{x} \in \mathcal{X}} \inf_{\bm{x}_j \in \bm{X}} \|\bm{x} - \bm{x}_j\|_2$ the \emph{fill distance}, $q_{\bm{X}} := \frac{1}{2} \min_{j \neq k} \|\bm{x}_j -\bm{x}_k\|_2$ the \emph{separation radius} and $\rho_{\bm{X},\mathcal{X}} := h_{\bm{X},\mathcal{X}}/q_{\bm{X}}$ the \emph{mesh ratio}. We will assume we evaluate all integrands on the same point set $\bm{X}$ which satisfies either of these assumptions:
\begin{enumerate}[label=(\subscript{A}{{\arabic*}})]
\item $\bm{X}$ consists of independently and identically distributed (IID) samples from some probability measure $\Pi'$ which admits a density $\pi'>0$ on $\mathcal{X}$.

\item $\bm{X}$ is a \emph{quasi-uniform} grid on $\mathcal{X} \subset \mathbb{R}^p$ (i.e. satisfies $h_{\bm{X},\mathcal{X}} \leq C_1 N^{-\frac{1}{p}}$ for some $C_1>0$) and satisfies $h_{\bm{X},\mathcal{X}} \leq C_2 q_{\bm{X},\mathcal{X}}$ for some $C_2>0$.
\end{enumerate}
Examples of point sets satisfying $(A_2)$ include uniform grid points in some hypercube.
\begin{theorem}[Convergence rate for BQ with separable kernel]\label{theorem:convergence_separable}
Suppose we want to approximate $\Pi[\bm{f}]$ for some $\bm{f}:\mathcal{X}\rightarrow \mathbb{R}^D$ and $\hat{\Pi}_{\text{BQ}}[\bm{f}]$ is the multi-output BQ rule with the kernel $\bm{C}(\bm{x},\bm{x}') = \bm{B} c(\bm{x},\bm{x}')$ for some positive definite $\bm{B} \in \mathbb{R}^{D \times D}$ and scalar-valued kernel $c:\mathcal{X}\times \mathcal{X}\rightarrow \mathbb{R}$. Then, $\forall d=1,\ldots,D$, we have:
\begin{eqnarray*}
e(\mathcal{H}_{\bm{C}},\hat{\Pi}_{\text{BQ}},\bm{X},d) 
& = &
\mathcal{O}\left( e(\mathcal{H}_c,\hat{\Pi}_{\text{BQ}},\bm{X} )\right).
\end{eqnarray*}
In particular, assume that $\mathcal{X} \subset \mathbb{R}^p$ satisfies an interior cone condition with Lipschitz boundary\footnote{Formally defined in Appendix \ref{appendix:additional_background_MOBQ} for completeness.} and $\bm{X}$ satisfies assumption $(A_1)$ or $(A_2)$. Then, the following rates hold:
\begin{itemize}
\item If $\mathcal{H}_c$ is norm-equivalent to an RKHS with Mat\'ern kernel of smoothness $\alpha>\frac{p}{2}$, we have $\forall d=1,\ldots,D$:
\begin{eqnarray*}
e(\mathcal{H}_{\bm{C}},\hat{\Pi}_{\text{BQ}},\bm{X},d) 
& = &
\mathcal{O}\left(N^{-\frac{\alpha}{p}+\epsilon}\right),
\end{eqnarray*}
for $\epsilon>0$ arbitrarily small.

\item If $\mathcal{H}_c$ is norm-equivalent to the RKHS with squared-exponential, multiquadric or inverse multiquadric kernel, we have $\forall d=1,\ldots,D$:
\begin{eqnarray*}
e(\mathcal{H}_{\bm{C}},\hat{\Pi}_{\text{BQ}},\bm{X},d) 
& = &
\mathcal{O}\left(\exp\left(-C_1 N^{\frac{1}{p}-\epsilon}\right)\right),
\end{eqnarray*}
for some $C_1>0$ and for some $\epsilon>0$ arbitrarily small.
\end{itemize}
\end{theorem}
\begin{proposition}[Convergence rate for sum of kernels]\label{cor:sum_separable_kernels}
Suppose that $\bm{C}(\bm{x},\bm{x}') = \sum_{q=1}^Q \bm{C}_q(\bm{x},\bm{x}')$. Then:
\begin{eqnarray*}
e(\mathcal{H}_{\bm{C}},\hat{\Pi}_{\text{BQ}},\bm{X},d) 
& = & \argmax_{q \in \{1,\ldots,Q\}}  \mathcal{O}\left( e(\mathcal{H}_{\bm{C}_q},\hat{\Pi}_{\text{BQ}},\bm{X},d) 
\right).
\end{eqnarray*}  
\end{proposition}
We clarify that the notation with $\epsilon$ is common in the numerical integration literature, and is used to hide powers of $\log n$ terms since these do not have a significant influence on the asymptotic convergence rate.

It is interesting to note that the rate of convergence for multi-output BQ is the same as that of uni-output BQ \citep{Briol2015a}. This can be explained intuitively by the fact that, when adding a new integrand, we can only gain by a constant factor since we always evaluate the functions at the same input points. In fact the proof of Thm. \ref{theorem:convergence_separable} provides an expression for this improvement factor (in terms of WCE) for any integrand $f_d$, and this depends explicitly on its correlation with the other functions: $|\sum_{i,j=1}^D (\bm{B}^{-1})_{ij} \bm{B}_{id} \bm{B}_{jd}|$. From a practitioner's viewpoint, this can clearly be used to balance the value of using several integrands with the additional computational cost incurred by using multi-output BQ.

We now give a result in the misspecified setting when the function $f$ is assumed to be smoother than it is. In this case, it is still possible to recover the optimal convergence rate: 
\begin{theorem}[Misspecified Convergence Result for Separable Kernel]\label{thm:misspecified_multioutputBQ} Let $c_{\alpha}$ be a kernel norm-equivalent to a Mat\'{e}rn kernel of smoothness $\alpha$ on some domain $\mathcal{X}$ with Lipschitz boundary and satisfying an interior cone condition. Consider the BQ rule $\hat{\Pi}_{\text{BQ}}[\bm{f}]$ corresponding to a separable kernel $\bm{C}_{\alpha}(x,x') = \bm{B} c_{\alpha}(x,x')$ with $\bm{X}$ satisfying $(A_2)$, and suppose that $\bm{f} \in \mathcal{H}_{\bm{C}_{\beta}}$ where $\frac{p}{2} \leq \beta \leq \alpha$. Then, $\forall d = 1,\ldots,D$:
\begin{eqnarray*}
\left| \Pi[f_d] - \hat{\Pi}_{\text{BQ}}[f_d]\right| & = & \mathcal{O}\left(N^{-\frac{\beta}{p}+\epsilon}\right),
\end{eqnarray*}
for some $\epsilon > 0$. 
\end{theorem}
This last theorem demonstrate that the method is rate adaptive as long as we choose a kernel which is too smooth. However, it also demonstrates a drawback of the separable kernels: if one of the integrands is rough but all other are smooth, then the worst-case error could potentially converge slowly for all of them.

Finally, we note that studying the method in other information complexity settings than the worst-case would also be interesting. For example, it is trivial to show that the method above satisfies the definition of Bayesian probabilistic numerical method of \cite{Cockayne2017} (Def. 2.5). Furthermore, optimality conditions for this method could also be obtained in a game-theoretic setting (in terms of a two-player mixed strategies game) by extending the theory on gamblets by \cite{Owhadi2017}.

\section{Applications}\label{sec:application}

\paragraph{Multi-fidelity modelling}
\begin{figure}[t!]
\begin{center}
\includegraphics[trim={5cm 0 0 0},clip,width=0.55\textwidth]{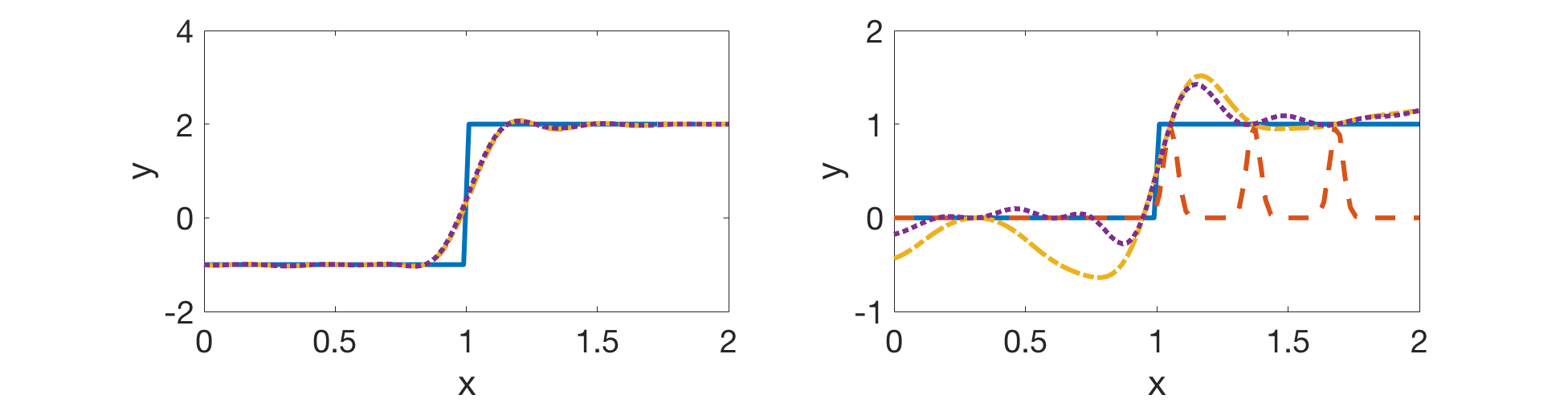}
\includegraphics[trim={5cm 0 0 0},clip,width=0.55\textwidth]{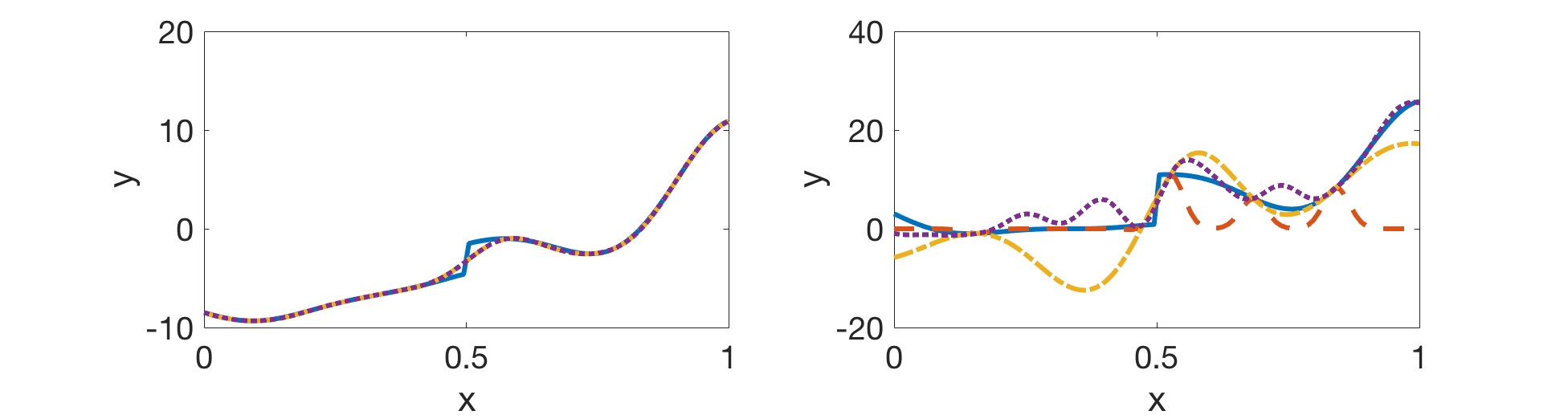}
\caption{Multi-fidelity modelling: Plot of the Step function (top), Forrester function (bottom) for the low fidelity (left) and high fidelity (right). Each plot gives the true function (blue) and their unit-output (dashed, red), LMC-based multi-output (dashed, yellow) and PC-based multi-output (dotted purple) approximations.
\vspace{-3mm}}
\label{plot:multifidelity_functions}
\end{center}
\end{figure}
Consider some function $f^{\text{high}}:\mathcal{X}\rightarrow \mathbb{R}$ representing some complex engineering model of interest, which we would like to use for some task such as statistical inference or optimization. These models usually require the simulation of underlying physical systems, which can make each evaluation prohibitively expensive and will therefore limit $N$ to the order of tens or hundreds. To tackle this issue, multi-fidelity modelling proposes to build cheap, but less accurate, alternatives $f^{\text{low}}_1,\ldots,f^{\text{low}}_{D-1}:\mathcal{X}\rightarrow \mathbb{R}$ to $f^{\text{high}}$, and use the cheaper models in order to accelerate computation for the task of interest. This can be done using surrogate models (e.g. support vector machines, GPs or neural networks), projection-based models (Krylov subspace or reduced basis methods) or models where the underlying physics is simplified; see \cite{Peherstorfer2016} for an overview.

In this section, we consider the problem of numerical integration in such a multi-fidelity setup. Note that two related methods for Monte Carlo estimation are the multi-fidelity Monte Carlo estimator \citep{Peherstorfer2016} and the multilevel Monte Carlo of \citep{Giles2015}, both of which are based on control variate identities. 

\begin{figure}[t!]
\begin{center}
\begin{tabular}{|c|c|c|c|} 
 \hline
 Model & BQ & LMC-BQ & PC-BQ \\  [0.5ex] 
 \hline\hline
 Step (l) & 0.02 (0.22) & 0.02 (0.21) & 0.02 (0.52) \\ 
 Step (h) & 0.41 (0.03) & 0.09 (0.09) & 0.04 (0.15)\\ 
 \hline
 For. (l) & 0.08 (4.91) & 0.08 (4.95) & 0.07 (33.95)\\ 
 For. (h) & 3.96 (3.98) & 2.86 (27.01) & 1.06 (63.80)\\ 
 \hline
\end{tabular}
\end{center}
\caption{Multi-fidelity modelling: Performance of uni-output BQ and multi-output BQ (with LMC and PC kernels) on the step function (Step) and the Forrester function with jump (For.) in the low fidelity (l) and high fidelity (h) cases. The values given are absolute errors with the posterior variance in brackets.}
\label{table:multifidelity}
\end{figure}
We approach this problem with multi-output BQ on the vector-valued function $\bm{f} = (f^{\text{high}},f^{\text{low}}_{1},\ldots,f^{\text{low}}_{D-1})^\top$. Note that multi-output Gaussian processes were already proposed for multi-fidelity modelling in 
\cite{Perdikaris2016,Parussini2017}, and we extend their methodologies to the task of numerical integration. We consider two toy problems from this literature \cite{Raissi2016} to highlight some of the advantages and disadvantages of our methodology
\begin{enumerate}
\item A \textit{step function} on $\mathcal{X} = [0,2]$:
\begin{equation*}
f_1^{\text{low}}(x) =  
\begin{cases}
0, \; x \leq 1\\
1, \; x > 1
\end{cases}
f^{\text{high}}(x) = 
\begin{cases}
-1, \; x \leq 1\\
2, \; x > 1
\end{cases}
\end{equation*}
\item The \textit{Forrester function with Jump} on $\mathcal{X} = [0,1]$:
\begin{equation*}
f_1^{\text{low}}(x) =  
\begin{cases}
\frac{(3x-1)^2 \sin(12x-4)}{4} +10(x-1), \; x \leq \frac{1}{2}\\
3+ \frac{(3x-1)^2\sin(12x-4)}{4} +10(x-1), \;  x > \frac{1}{2}
\end{cases} 
\end{equation*}
\begin{equation*}
f^{\text{high}}(x) = 
\begin{cases}
\quad 2 f^{\text{low}}(x)-20(x-1), \; x \leq \frac{1}{2}\\
4+2 f^{\text{low}}(x)-20(x-1), \;  x > \frac{1}{2}
\end{cases}
\end{equation*}
\end{enumerate}

The functions and conditioned GPs are given in Fig. \ref{plot:multifidelity_functions}, whilst the uni-output and multi-output BQ estimates for integration of these functions against a uniform measure $\Pi$ are given in the table in Fig. \ref{table:multifidelity}. In both cases, $20$ equidistant points are used, with point number $4,10,11,14$ and $17$ used to evaluate the high fidelity model and the others used for the low fidelity model. The choice of kernel hyperparameters is made by maximising the marginal likelihood (often called empirical Bayes). Further details, and an additional test function can be found in Appendix \ref{sec:appendix_multifidelity}.

Note that both of these problems are challenging for several reasons. Firstly, due to their discontinuity, the integrands are not in the RKHS $\mathcal{H}_{\bm{C}}$ corresponding to the kernel $\bm{C}$ used in multi-output BQ. In particular, the problems are misspecified in the sense that the true function is not in the support of the prior. It is therefore difficult to interpret the posterior distribution on $\Pi[\bm{f}]$, and we end up with credible intervals which are too wide. This is for example illustrated in the values of the posterior variance for the high-fidelity Forrester function. Secondly, in each case, the high and low-fidelity models are defined on different scales and so require tuning of several kernel hyper-parameters. This can of course make it challenging for multi-output BQ since the number of function evaluations $N$ is small and empirical Bayes will tend to be inefficient in those cases.

However, despite these two issues, it is interesting to note that both of the multi-output BQ methods manage to significantly outperform uni-output BQ in terms of point estimate, as the sharing of information allows the multi-output models to better represent the main trends in the functions. Furthermore, the multi-output BQ does not suffer from the issues of overconfident posterior credible intervals present in uni-output BQ; contrast for example the posterior variances for the high-fidelity step function.


\paragraph{Global illumination}

In this section, we apply multi-output BQ to a challenging numerical integration problem from the field of computer graphics, known as global illumination. BQ was previously applied to this problem in several papers \cite{Brouillat2009,Marques2013,Briol2015a}, but we propose to extend these results using multi-output BQ.

Global illumination is a problem which occurs when trying to obtain realistic representation of light interactions for the design of virtual environments (e.g. a video game). One model of the amount of light coming from an object towards the camera (representing the current viewpoint on this environment) is given by the following equation: 
\begin{eqnarray*}
L_0(\omega_0) & = & L_e(\omega_0) + \int_{\mathbb{S}^2} L_i(\omega_i) \rho(\omega_i,\omega_0) [\omega_i \cdot n]_{+} \mathrm{d}\Pi(\omega_i).
\end{eqnarray*}
where $[x]_+ = \text{max}(0,x)$. The function $L_0:\mathbb{S}^2 \rightarrow \mathbb{R}$ evaluated at $\omega_0$ is called the outgoing radiance in direction $\omega_0$ (the angle of the outgoing light from the object normal $n$), $L_e(\omega_0):\mathbb{S}^2 \rightarrow \mathbb{R}$ is the amount of light emitted by the object, and $L_i:\mathbb{S}^2 \rightarrow \mathbb{R}$ evaluated at $\omega_i$ is the amount of light reflected by the object (which originated from an angle $\omega_i$ from the object's normal $n$). Here, $\mathbb{S}^2 = \{\bm{x} = (x_1,x_2,x_3) \in \mathbb{R}^3 : \|\bm{x}\|_{2} = 1\}$ and $\rho(\omega_i,\omega_0):\mathbb{S}^2 \times \mathbb{S}^2 \rightarrow \mathbb{R}$ is called the bidirectional reflectance distribution and represents the proportion of light being reflected. 

We follow \cite{Briol2015a} and consider the problem as $\Pi[h^{\omega_0}] = \int_{\mathbb{S}^2} h^{\omega_0}(\omega_i) \Pi(\mathrm{d}\omega_i)$ where $\Pi$ is the uniform measure on $\mathbb{S}^2$, and $h^{\omega_0}(\omega_i) = L_i(\omega_i)\rho(\omega_i,\omega_0)[\omega_i\cdot \omega_0]_{+}$ is a function which can be evaluated by making a call to an environment map (which we consider to be a black box). One scenario which is common in these type of problems is to look at an object from different angles $\omega_0$, with the camera moving. In this case, it is reasonable to assume that the different integrands $h^{\omega_0}$ will be very similar when the difference in the angle $\omega_0$ is small, and it is therefore natural to consider jointly estimating their integrals. In the experiments we consider five integrands $f_i = h^{\omega_0^i}$ for $i=1,\ldots,5$ where $\omega_0^1,\ldots,\omega_0^5$ are on a great circle of the sphere at intervals determined by an angle of $0.005 \pi$. 

We therefore consider two-output and five-output BQ with independent and identically distributed (Monte Carlo) samples $\bm{X}$ from the uniform measure $\Pi$. We propose to use a separable kernel with scalar-valued RKHS $\mathcal{H}_c$ being a Sobolev space of smoothness $\frac{3}{2}$ over $\mathbb{S}^2$ and has kernel $c(\bm{x},\bm{x}') = \frac{8}{3}-\|\bm{x}-\bm{x}'\|_2^2$. For the matrix $B$ representing the covariance between outputs, we propose to make this covariance proportional to the difference in angle at which the camera looks at the object. In particular we choose $(B)_{ij} = \exp(\omega_0^i \cdot \omega_0^j -1)$ for simplicity, but this could be generalised to include a lengthscale and amplitude hyperparameter to be learnt together with the hyperparameters of the scalar-valued kernel $c$. 

\begin{figure}[t!]
\begin{center}
\includegraphics[trim={2cm 0 2cm 0},clip,width=0.23\textwidth]{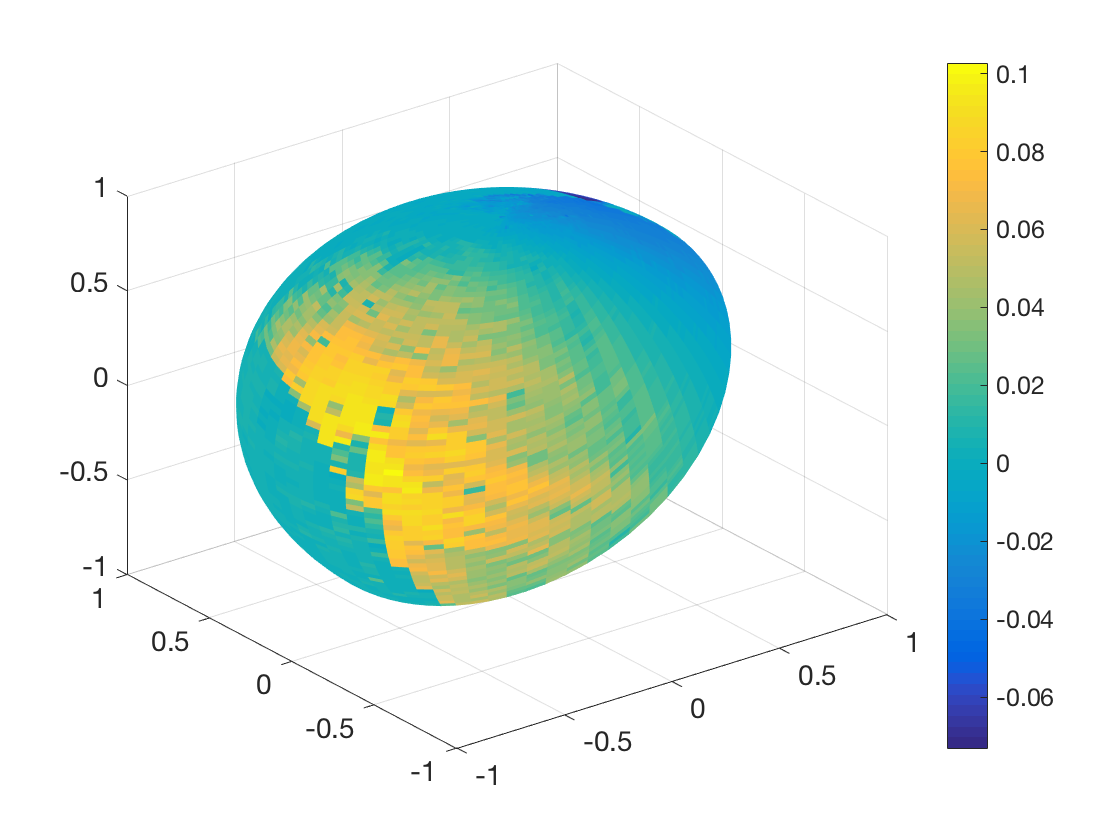}
\includegraphics[trim={2cm 0 2cm 0},clip,width=0.23\textwidth]{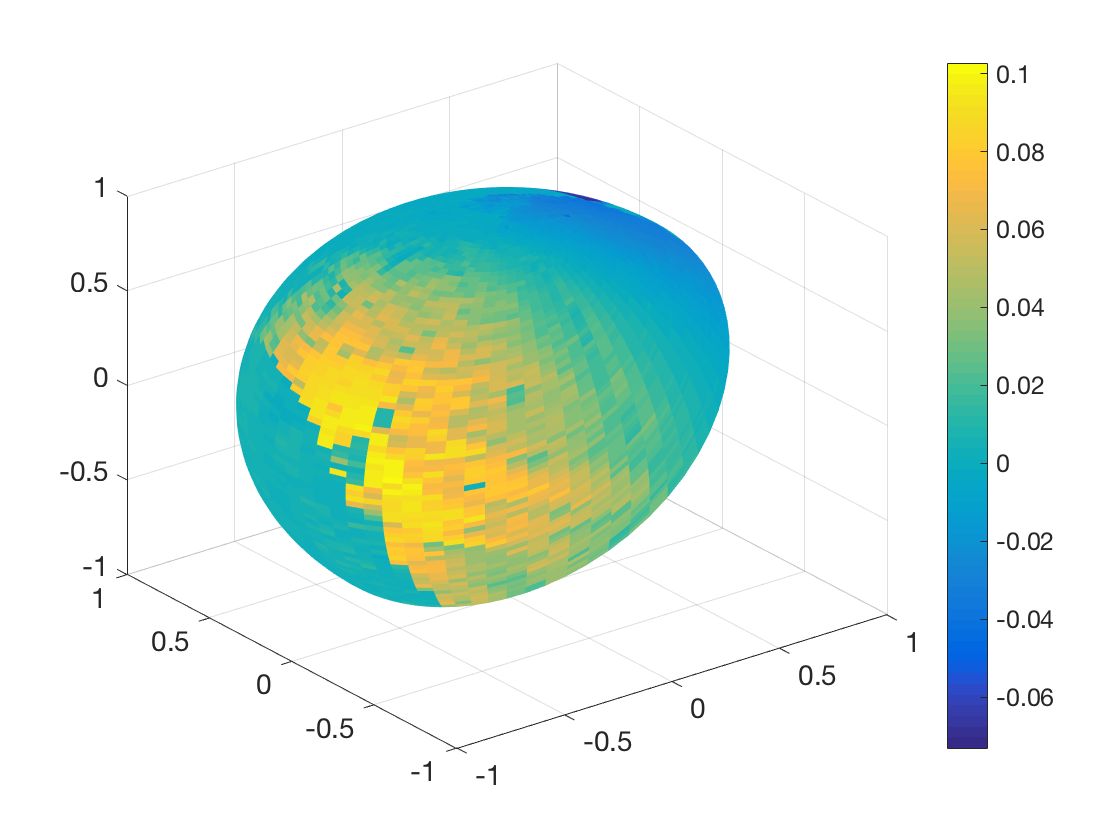} \\
\includegraphics[trim={2cm 0 2cm 0},clip,width=0.23\textwidth]{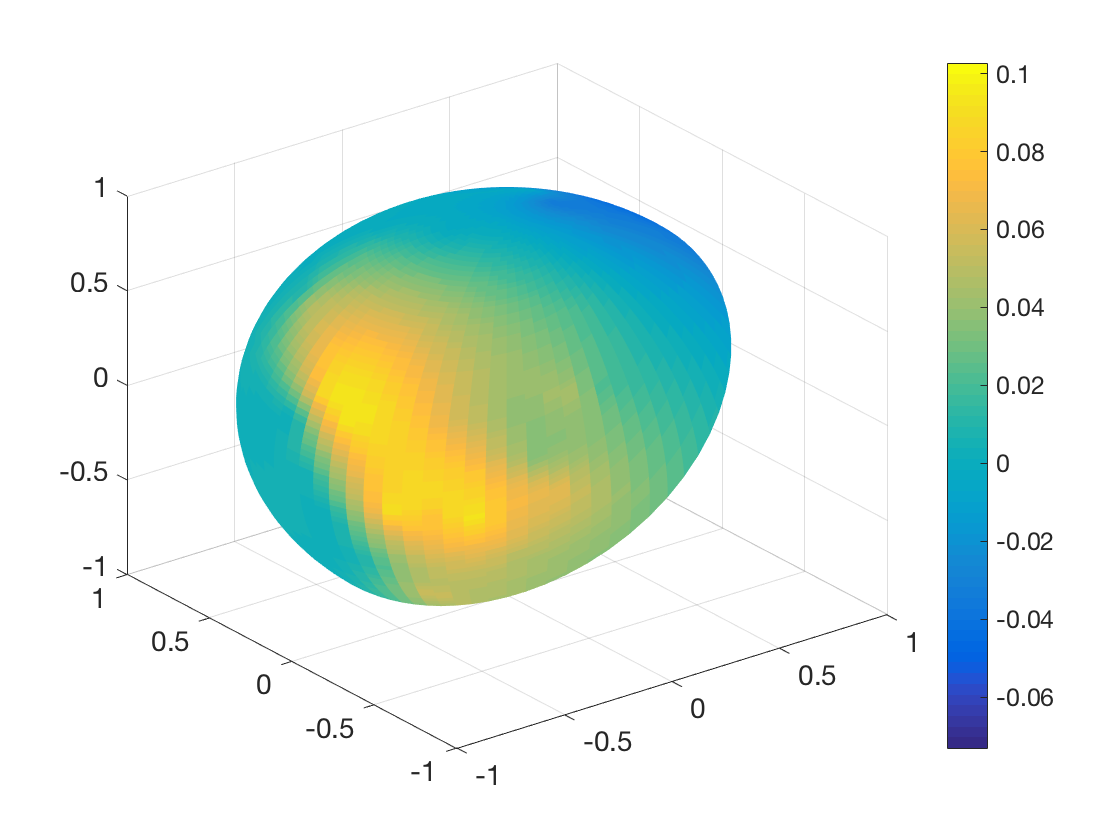}
\includegraphics[trim={2cm 0 2cm 0},clip,width=0.23\textwidth]{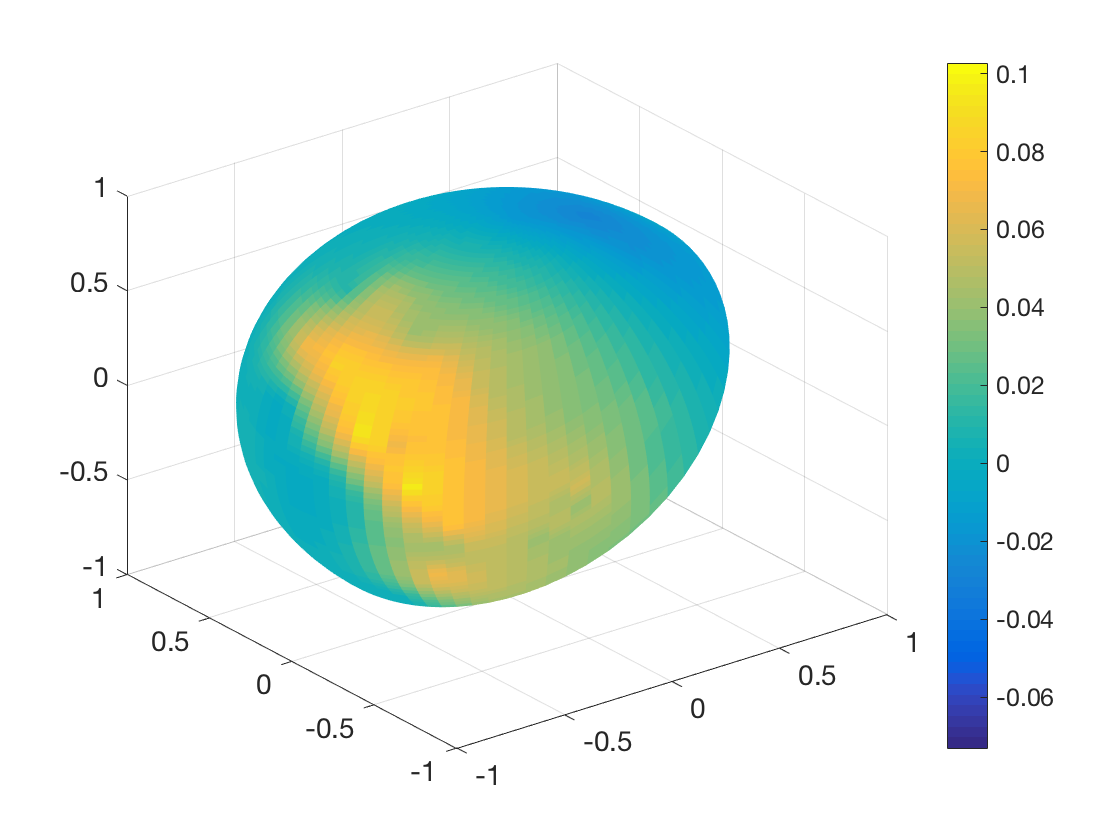}\\
\includegraphics[trim={2cm 0 2cm 0},clip,width=0.23\textwidth]{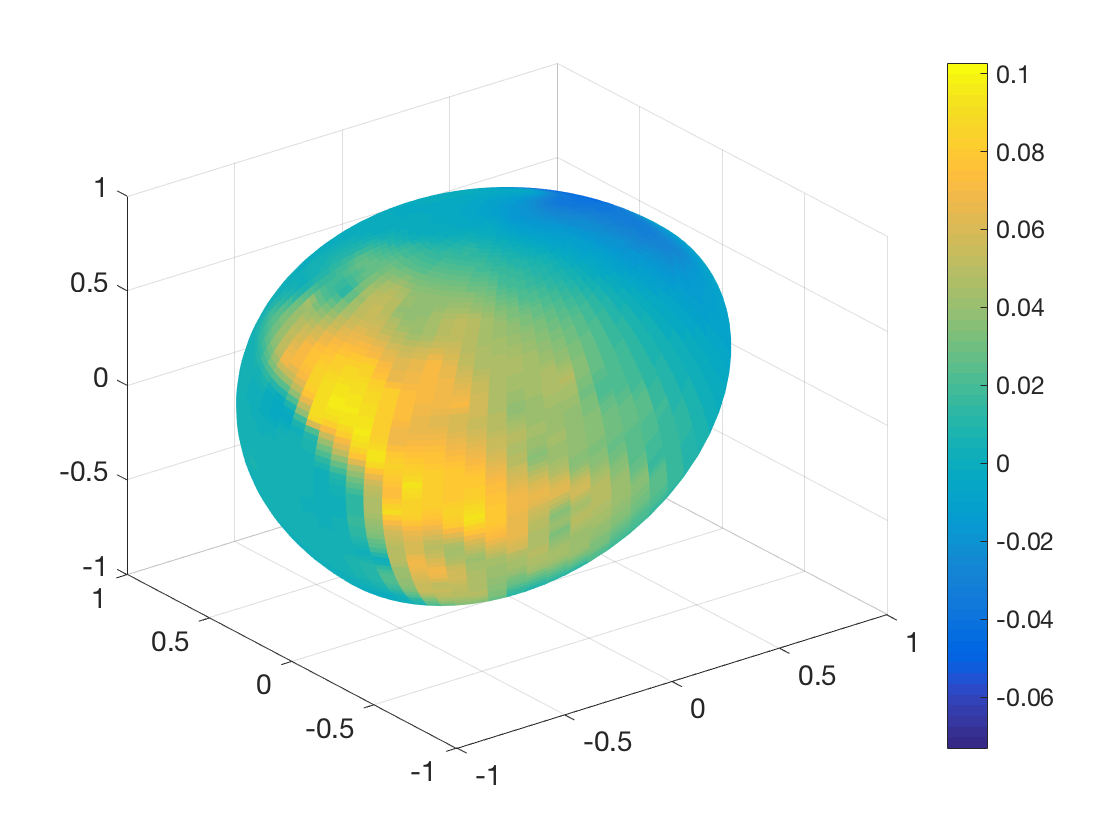}
\includegraphics[trim={2cm 0 2cm 0},clip,width=0.23\textwidth]{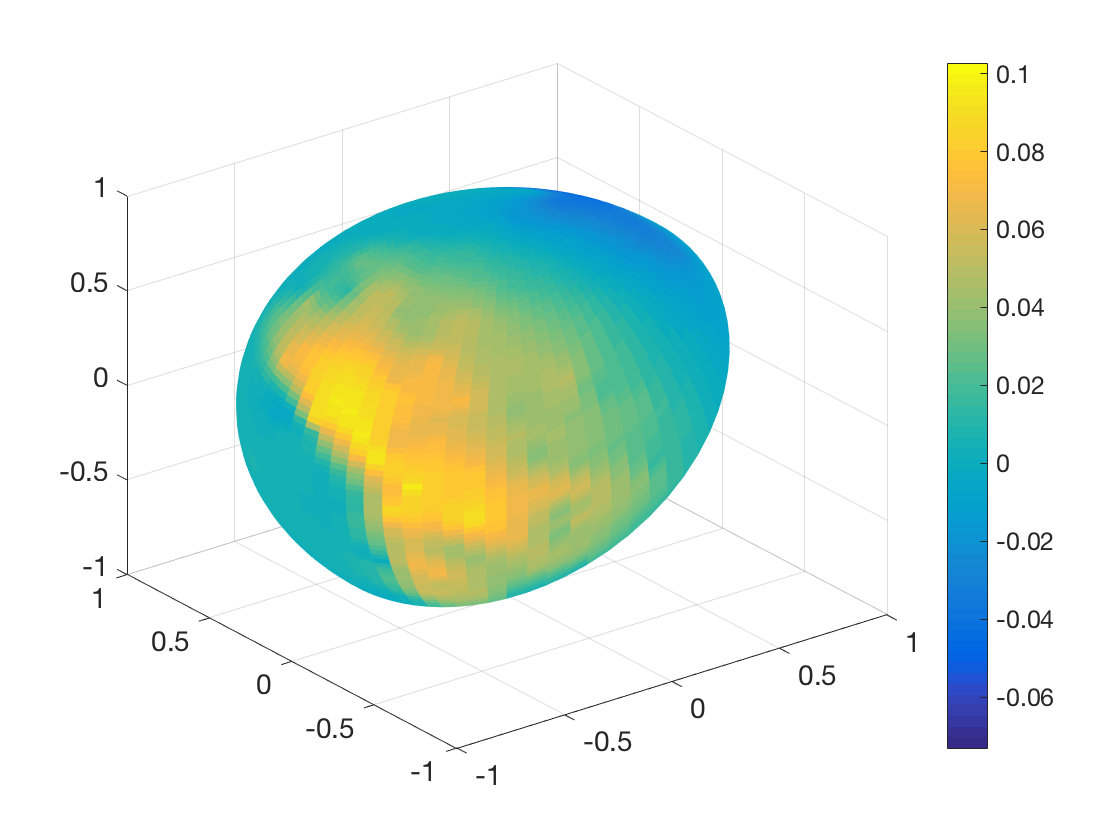}\\
\caption{Global illumination: Plot of $f_1$ (left) and $f_2$ (right) in the case of the red channel. The plots give the true functions being approximated (top), the uni-output approximations (middle) and the five-output approximations with LMC kernel (bottom).
\vspace{-3mm}}
\label{fig:sphere_plots}
\end{center}
\end{figure}

\begin{figure}[t!]
\begin{center}
\includegraphics[trim={2cm 0 1cm 0},clip,width=0.495\textwidth]{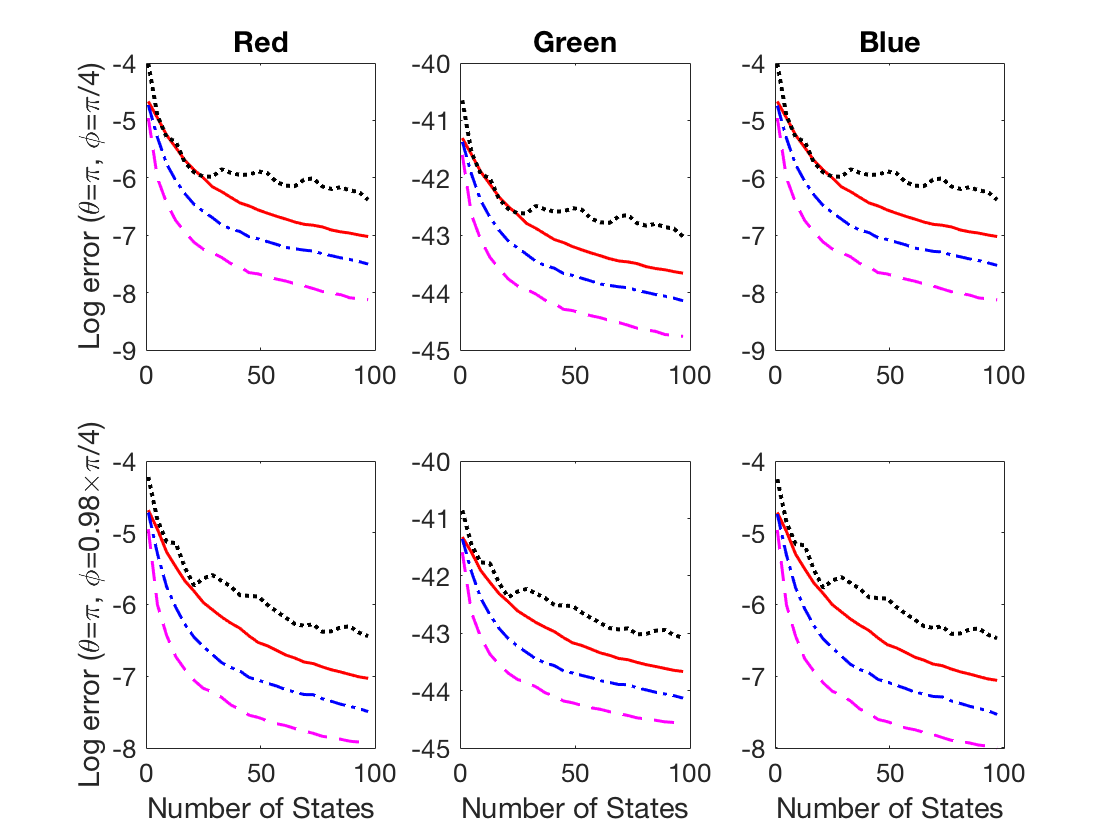}
\caption{Global illumination: Plot of error estimates for $f_1$ (top) and $f_2$ (bottom), in the case of the red, green and blue channels. The log-error is plotted for uni-output BQ (red), two-output BQ based on LMC (blue), five-output BQ based on LMC (magenta) and standard Monte Carlo (dotted black).
\vspace{-3mm}}
\label{plot:global_illum_error}
\end{center}
\end{figure}

The GP means for the one-output and five-output cases are given in Fig. \ref{fig:sphere_plots}, and we can clearly notice a significant improvement in approximation accuracy with the larger number of outputs.
Results for integration error are given in Fig. \ref{plot:global_illum_error}. As noticed, the integration error (for a fixed number of evaluations $N$ of each integrand) is significantly reduced by increasing the number of outputs $D$. The individual posterior variances for this problem (see Appendix \ref{sec:appendix_global_illum} Fig. \ref{plot:global_illum_WCE}) are also smaller, reflecting the fact that our uncertainty is reduced due to use of observations from other integrands. 

In fact, a small extension of Thm. \ref{theorem:convergence_separable} (combined with the rate for the scalar-valued kernel in \cite{Briol2015a}) allows us to obtain an asymptotic convergence rate for the posterior variance on each integral $\Pi[f_d]$:
\begin{corollary}
Let $\mathcal{X}$ be the sphere $\mathbb{S}^2$ and $\bm{X}$ be IID uniform points on $\mathcal{X}$. Assume $\bm{C}$ is a separable kernel with $c$ defined above. Then $e(\mathcal{H}_{\bm{C}},\hat{\Pi}_{\text{BQ}},\bm{X},d)  =  \mathcal{O}_P \left(N^{-\frac{3}{4}}\right)$.
\end{corollary}
The same rate with improved rate constant was observed in \citep{Briol2015a} when using QMC point sets, and similar gains could be obtained in this multi-output case.

We note that there a significant potential further gains for the use of multi-output BQ in this setting. Similar integration problems need to be computed for three colors in every pixel of an image, and for every image in a video. This is challenging computationally and limits the use of Monte Carlo methods to a few dozen points. Designing specific matrix-valued kernels could provide enormous gains since we end up with thousands of correlated integrands. Furthermore, the weights only depend on the choice of kernel and not on function values, so that all of the weights could be pre-computed off-line to be later used in real-time.

\section{Conclusion}
\label{sec:conclusion}
 
We have proposed an extension of Bayesian Quadrature to the case where we are interested in numerically computing the integral of several functions which are related. In particular we have proposed a new algorithm based on jointly modelling the integrands with a Gaussian prior. Then, we provided a theoretical study of the rate of convergence for the case where the kernel is separable and illustrated the potential of our methodology on applications in multi-fidelity modelling and computer graphics. Our main contribution however, has been to highlight the natural extension of Bayesian probabilistic numerical methods to the joint estimation of the solution of several numerical problems (in this case, numerical integration problems).

There are several possible extensions of multi-output BQ which we reserve for future work. One important question remaining is that of the choice of sampling distribution. In the uni-output case, it is well known that obtaining an optimal sampling distribution with respect to the $\mathbb{V}_n[\Pi[f]]$ is intractable in most cases. \cite{Briol2017} proposed an algorithm to approach such a distribution, and \cite{Kanagawa2017} provided conditions on the point sets to guarantee fast convergence. In the multi-output case, the problem is even more complex due to the interaction between the different integration problems. However, the literature on the design of experiments for co-kriging/multi-output GPs may be of interest, and the use of more advanced sampling distributions will certainly provide significant gains.

%
\subsection*{Acknowledgements} 
The authors are grateful to Alessandro Barp, Aretha Teckentrup, Chris Oates and Motonobu Kanagawa for helpful discussions. FXB was supported by the EPSRC grants [EP/L016710/1, EP/R018413/1, EP/N510129/1]. MG was supported by the EPSRC grants [EP/J016934/3, EP/K034154/1, EP/P020720/1, EP/R018413/1, EP/N510129/1], an EPSRC Established Career Fellowship, the EU grant [EU/259348] and the Lloyds Register Foundation Programme on Data-Centric Engineering. The authors would like to thank the Isaac Newton Institute for Mathematical Sciences for support and hospitality during the programme on ``Uncertainty Quantification for Complex Systems: Theory and Methodologies". This work was supported by EPSRC grant no [EP/K032208/1]. Finally, this material was also based upon work partially supported by the National Science Foundation under Grant DMS-1127914 to the Statistical and Applied Mathematical Sciences Institute. 


\bibliography{VectorPI}
\bibliographystyle{icml2018}

\newpage
\onecolumn
\appendix

\section{Additional Background}\label{appendix:additional_background_MOBQ}

\subsection{Sobolev Spaces}

The RKHS induced from the Mat\'{e}rn kernel $c_\alpha$ defined in Equation \ref{eq:matern_kernels} is norm-equivalent to a Sobolev space \citep{Adams2003}. When $\alpha \in \mathbb{N}$, these spaces are defined as:
\begin{eqnarray*}
W_2^{\alpha}(\mathcal{X}) & := & \left\{ f \in L_2(\mathcal{X}) : D^{\nu}f \in L_2(\mathcal{X})\text{  exists } \forall \nu \in \mathbb{N}^p_0 \text{ with } |\nu | \leq \alpha \right\},
\end{eqnarray*}
with inner product 
\begin{eqnarray*}
\langle f,g \rangle_{W_2^\alpha(\mathcal{X})} & := & \sum_{|\nu| \leq \alpha} \langle D^\nu f , D^\nu g\rangle_{L_2(\mathcal{X})}
\end{eqnarray*}
for all $f,g \in W_2^\alpha(\mathcal{X})$. This means that all functions in $\mathcal{H}_{k_{\alpha}}$ will have smoothness $\alpha$ (here $D^{\nu}$ denotes the total derivative corresponding to the multi-index $\nu  = (\nu_1,\ldots,\nu_p) \in \mathbb{N}_0^p$).

It is also possible to have fractional Sobolev spaces; i.e. the smoothness $\alpha>0$ can take any positive real value. For $\mathcal{X} = \mathbb{R}^p$ and denoting by $\hat{f}$ the Fourier transform of $f$, these spaces are given by:
\begin{eqnarray*}
H^\alpha(\mathbb{R}^p) & := & \left\{ f \in L_2(\mathbb{R}^p) :  \int |\hat{f}(\xi) |^2 (1 + \|\xi\|^2)^\alpha \mathrm{d}\xi < \infty \right\}
\end{eqnarray*}
with associated inner product:
\begin{eqnarray*}
\left\langle f,g \right\rangle_{H^\alpha(\mathbb{R}^p)} & := & \int \hat{f}(\xi) \overline{\hat{g}(\xi)} (1 + \| \xi \|^2)^{\alpha} \mathrm{d}\xi 
\end{eqnarray*}
 for all $f,g \in H^\alpha(\mathbb{R}^d)$ where $\overline{\hat{g}}$ denoted the complex conjugate of $\hat{g}$.

\subsection{Lipschitz Boundary Conditions}

In this section we introduce the notion of Lipschitz boundary condition, which is required for our domain $\mathcal{X}$ in the theory in Section \ref{sec:MOBQ_theory}. The introduction in this section follows that of Section 3 in \cite{Kanagawa2017}.

To do so, we begin by introducing \emph{special Lispchitz domains}. For $d>2$, we say that an open set $\mathcal{X} \subset \mathbb{R}^p$ is a special Lipschitz domain if there exists a rotation of $\mathcal{X}$, denoted by $\tilde{\mathcal{X}}$, and a function $\phi:\mathbb{R}^{p-1} \rightarrow \mathbb{R}$ that satisfy the following:
\begin{enumerate}
\item $\tilde{\mathcal{X}} = \{(x,y) \in \mathbb{R}^p: y>\phi(x)\}$.

\item $\phi$ is a Lipschitz function such that $|\phi(x) - \phi(x')| \leq M \|x-x'\| \forall x,x' \in \mathbb{R}^{p-1}$, where $M>0$ is called the Lipschitz bound of $\mathcal{X}$.
\end{enumerate}
With this definition now complete, we can define the notion of a domain with Lipschitz boundary. Let $\mathcal{X} \subset \mathbb{R}^p$ be an open set and $\partial \mathcal{X}$ be its boundary. We say the boundary is Lipschitz $\exists \epsilon,M >0, K \in \mathbb{N}$ and open sets $U_1,\ldots,U_L \subset \mathbb{R}^p$ where $L \in \mathbb{N} \cup \{\infty\}$ such that the following holds:
\begin{enumerate}
\item For any $x \in \partial \mathcal{X}$, $\exists i$ such that $B(x,\epsilon)$, the ball centred at $x$ of radius $\epsilon$, satisfies $B(x,\epsilon) \subset U_i$.  

\item $U_{i_1} \cap \ldots \cap U_{i_{K+1}} = \emptyset$ for any distinct indices $\{i_{1},\ldots,i_{K+1}\}$.

\item For each index $i$, $\exists$ a special Lipschitz domain $\mathcal{X}_i \subset \mathbb{R}^p$ with Lipschitz bound $b$ such that $U_i \cap \mathcal{X} = U_i \cap \mathcal{X}_i$ and $b \leq M$.
\end{enumerate}

\section{Proofs}\label{sec:appendix_proofs}

\subsubsection*{Proof of Proposition \ref{prop:multioutput_BQ}}
\begin{proof} This proof follows directly the proof for the uni-output case in \cite{Briol2015a}. Suppose we have a prior on $\bm{f}$, denoted $\bm{g}$, which is a Gaussian process $\mathcal{GP}(\bm{0},\bm{C})$. Conditioning on some observations $(\bm{X},\bm{Y})=\{(\bm{X}_j,\bm{Y}_j)\}_{j=1}^D$, we get a Gaussian process posterior $\bm{g}_N$ where the mean and covariance functions are given by:
\begin{eqnarray*}
\bm{m}_N(\bm{x}) & = & \bm{C}(\bm{x},\bm{X})\bm{C}(\bm{X},\bm{X})^{-1}\bm{f}(\bm{X}),\\
\bm{C}_N(\bm{x},\bm{x}') & = & \bm{C}(\bm{x},\bm{x}') - \bm{C}(\bm{x},\bm{X}) \bm{C}(\bm{X},\bm{X})^{-1} \bm{C}(\bm{X},\bm{x}').
\end{eqnarray*}
Several applications of Fubini's theorem on each element of the vectors give:
\begin{eqnarray*}
\mathbb{E} \left[\Pi[\bm{g}_N]\right] 
& = & 
\int_{\Omega} \int_{\mathcal{X}} \bm{g}_N(\bm{x},\omega) \Pi(\mathrm{d}\bm{x})\mathbb{P}(\mathrm{d}\omega)
\; = \;
\int_{\mathcal{X}} \bm{m}_N(\bm{x}) \Pi(\mathrm{d}\bm{x}) \; = \; 
\Pi[\bm{C}(\cdot,\bm{X})]\bm{C}(\bm{X},\bm{X})^{-1}\bm{f}(\bm{X}), \\
\mathbb{V}[\Pi[\bm{g}_N]] 
& = & 
\int_{\Omega} \left[\int_{\mathcal{X}} \bm{g}_N(\bm{x},\omega) \Pi(\mathrm{d}\bm{x}) - \int_{\mathcal{X}} \bm{m}_N(\bm{x}) \Pi(\mathrm{d}\bm{x}) \right]^2 \mathbb{P}(\mathrm{d}\omega)\\
& = & \int_{\mathcal{X}} \int_{\mathcal{X}} \int_{\Omega} \left[\bm{g}_N(\bm{x},\omega) - \bm{m}_N(\bm{x})\right] \left[\bm{g}_N(\bm{x}',\omega) - \bm{m}_N(\bm{x}')\right] \mathbb{P}(\mathrm{d}\omega) \Pi(\mathrm{d}\bm{x}) \Pi(\mathrm{d}\bm{x}')\\
& = & \int_{\mathcal{X}} \int_{\mathcal{X}} \bm{C}_N(\bm{x},\bm{x}') \Pi(\mathrm{d}\bm{x}) \Pi(\mathrm{d}\bm{x}') \\
& = & 
\Pi\bar{\Pi}[\bm{C}] - \Pi[\bm{C}(\cdot,\bm{X})] \bm{C}(\bm{X},\bm{X})^{-1} \bar{\Pi}[\bm{C}(\bm{X},\cdot)].
\end{eqnarray*} 

\end{proof}

\subsubsection*{Proof of Proposition \ref{prop:optimal_weights}}

\begin{proof}
Denote by $\bm{e}_d$ the vertical vector of length $D$ with $d^{\text{th}}$ entry taking value $1$ and all other entries taking value $0$, and by $\bm{C}^d_{\bm{x}}(\bm{y}) = \bm{C}(\bm{y},\bm{x})\bm{e}_d$ the $d^{\text{th}}$ column of $\bm{C}(\bm{y},\bm{x})$. We notice that the representer of the integral is given by:
\begin{eqnarray*}
\Pi[f_d] 
& = & \Pi[\bm{f}^\top \bm{e}_d] 
\; = \; \Pi\left[\left\langle \bm{f},\bm{C}(\cdot,\bm{x})\bm{e}_d\right\rangle_{\bm{C}}\right] 
\; = \;
\left\langle \bm{f}, \Pi\left[\bm{C}(\cdot,\bm{x})\bm{e}_d \right] \right\rangle_{\bm{C}} 
\; = \;
\left\langle \bm{f}, \Pi\left[\bm{C}^d_{\bm{x}}\right] \right\rangle_{\bm{C}}.
\end{eqnarray*}
Using the Cauchy-Schwartz inequality, we get:
\begin{eqnarray*}
\left| \Pi[f_d] - \hat{\Pi}[f_d]\right| & \leq & \|\bm{f}\|_{\bm{C}} \left\| \Pi[\bm{C}^d_{\bm{x}}] - \hat{\Pi}[\bm{C}^d_{\bm{x}}] \right\|_{\bm{C}}.
\end{eqnarray*}
Taking supremums, we then obtain the following expression for the worst-case integration error:
\begin{eqnarray*}
\sup_{\|f\|_{\bm{C}}\leq 1} \left| \Pi[f_d] - \hat{\Pi}[f_d]\right| & = & \left\| \Pi[\bm{C}^d_{\bm{x}}] - \hat{\Pi}[\bm{C}^d_{\bm{x}}] \right\|_{\bm{C}}.
\end{eqnarray*}
We note that $\Pi[\bm{C}^d_{\bm{x}}] \in \mathcal{H}_{\bm{C}}$ and that the multi-output BQ rule is given by
\begin{eqnarray*}
\hat{\Pi}_{\text{BQ}}[\bm{C}^d_{\bm{x}}] & = & \Pi[\bm{C}(\cdot,\bm{X})]\bm{C}(\bm{X},\bm{X})^{-1}\bm{C}^d_{\bm{x}}(\bm{X}),
\end{eqnarray*}
and corresponds to an optimal interpolant in the sense of Thm 3.1 \cite{Micchelli2008}. We must therefore have that, for fixed quadrature points $\bm{X}$, any  quadrature rule $\hat{\Pi}[\bm{C}^d_{\bm{x}}]$ satisfies:
\begin{eqnarray*}
\left\| \Pi[\bm{C}^d_{\bm{x}}] - \hat{\Pi}_{\text{BQ}}[\bm{C}^d_{\bm{x}}] \right\|_{\bm{C}} 
& \leq &
\left\| \Pi[\bm{C}^d_{\bm{x}}] - \hat{\Pi}[\bm{C}^d_{\bm{x}}] \right\|_{\bm{C}}.
\end{eqnarray*}
Combining the equation above with the expression for the worst-case integration error of $f_d$ gives us our final result.

\end{proof}

\subsubsection*{Proof of Theorem \ref{theorem:convergence_separable}}
\begin{proof}
For the sake of clarity, we will distinguish between uni-output BQ and multi-output BQ rules and weights by adding subscripts corresponding to their kernel; i.e. $\Pi_{\text{BQ}}^{\bm{C}}[f]$ and $\bm{W}_{\text{BQ}}^{\bm{C}}$ denote the multi-output case and $\Pi_{\text{BQ}}^{c}[f]$ and $\bm{W}_{\text{BQ}}^{c}$ denote the uni-output case.

We start this proof by writing an expression for the weights of the multi-output BQ algorithm in terms of weights for the uni-output BQ algorithm:
\begin{eqnarray*}
\bm{W}_{\text{BQ}}^{\bm{C}} & = & \left(\Pi[\bm{C}(\cdot,\bm{X})] \bm{C}(\bm{X},\bm{X})^{-1}\right)^\top \\
& = & \left(\left( \Pi[\bm{B} \otimes \bm{c}(\cdot ,\bm{X})] \right)\left(\bm{B} \otimes c(\bm{X},\bm{X})\right)^{-1}\right)^\top\\
& = & \left(\left( \bm{B} \otimes \Pi[\bm{c}(\cdot ,\bm{X})] \right)\left(\bm{B}^{-1} \otimes c(\bm{X},\bm{X})^{-1}\right)\right)^\top\\
& = &  \left(\bm{B}\bm{B}^{-1}  \otimes \Pi[\bm{c}(\cdot ,\bm{X})]c(\bm{X},\bm{X})^{-1}\right)^\top\\
& = & \left(\bm{I}_D \otimes \left(\bm{w}_{\text{BQ}}^{c}\right)^\top\right)^\top \\
& = & \bm{I}_D \otimes \bm{w}_{\text{BQ}}^{c}.
\end{eqnarray*}
Using the above, we can find an expression for the multi-output BQ approximation with some kernel $\bm{C}_1 = \bm{B} c_1$ of the project mean element with respect to kernel $\bm{C}_2 = \bm{B} c_2$ in terms of the uni-output BQ approximation with kernel $c_1$ of the kernel mean of $c_2$. 
\begin{equation*}
\begin{split}
\hat{\Pi}^{\bm{C}_1}_{\text{BQ}}[(\bm{C}_2)^{d}_{\bm{x}}] 
& \; = \; \left(\bm{W}_{\text{BQ}}^{C_1}\right)^\top(\bm{C}_2)_{\bm{x}}^d(\bm{X}) \\
& \; = \; (I \otimes \bm{w}_{\text{BQ}}^{c_1})^\top (\bm{C}_2)_{\bm{x}}^d(\bm{X}) \\
& \; = \; (\bm{I} \otimes \bm{w}^{c_1}_{\text{BQ}})^\top (\bm{B} \bm{e}_d \otimes c_2(\bm{X},\bm{x})) \\
& \; = \; \bm{I} \bm{B} \bm{e}_d  \otimes \left(\bm{w}^{c_1}_{\text{BQ}}\right)^\top c_2(\bm{X},\bm{x})  \\
& \; = \; \bm{B} \bm{e}_d \hat{\Pi}^{c_1}_{\text{BQ}}[c_2(\cdot,\bm{x})].
\end{split}
\end{equation*}
As discussed, taking both kernels to be the same, the integration error for each individual integrand can be bounded as follows:
\begin{equation*}
\begin{split}
\sup_{\|f\|_{\bm{C}_2} \leq 1} \left| \Pi[f_d] - \hat{\Pi}^{\bm{C}_1}_{\text{BQ}}[f_d]\right|^2 
& \; = \; \left\| \Pi\left[(\bm{C}_2)^d_{\bm{x}}\right] - \hat{\Pi}^{\bm{C}_1}_{\text{BQ}}\left[(\bm{C}_2)^d_{\bm{x}}\right]\right\|^2_{\bm{C}_2} \\
& \; = \; \left\|  (\bm{B} \bm{e}_d) \left(\Pi\left[c_2(\cdot,\bm{x})\right] - \hat{\Pi}^{c_1}_{\text{BQ}}\left[c_2(\cdot,\bm{x})\right]\right)\right\|^2_{\bm{C}_2}\\
& \; = \; \sum_{i,j=1}^D (\bm{B}^{-1})_{ij} \times  \Big\langle \bm{B}_{id} (\Pi\left[c_2(\cdot,\bm{x})\right] - \hat{\Pi}^{c_1}_{\text{BQ}}\left[c_2(\cdot,\bm{x})\right]) ,  \bm{B}_{jd} (\Pi\left[c_2(\cdot,\bm{x})\right] - \hat{\Pi}^{c_1}_{\text{BQ}}\left[c_2(\cdot,\bm{x})\right]) \Big\rangle_{c_2}\\
& \; = \;  \sum_{i,j=1}^D (\bm{B}^{-1})_{ij} \bm{B}_{id} \bm{B}_{jd}  \left\|\Pi\left[c_2(\cdot,\bm{x})\right] - \hat{\Pi}^{c_1}_{\text{BQ}}\left[c_2(\cdot,\bm{x})\right]\right\|_{c_2}^2  \\
& \; \leq \; K \left\|\Pi\left[c_2(\cdot,\bm{x})\right] - \hat{\Pi}^{c_1}_{\text{BQ}}\left[c_2(\cdot,\bm{x})\right]\right\|_{c_2}^2.
\end{split}
\end{equation*}
Here, we first used the definition of worst-case error, then the definition of the $H_{\bm{C}_2}$ norm in terms of $\mathcal{H}_{c_2}$ norm (as given for the seperable kernel in \cite{Alvarez2011}), and the final inequality follows by taking $K>0$ to be $K = |\sum_{i,j=1}^D (\bm{B}^{-1})_{ij} \bm{B}_{id} \bm{B}_{jd}|$. 
Taking the square-root on either side gives us:
\begin{eqnarray}\label{eq:thm1_proof_step}
\sup_{\|f\|_{\bm{C_2}} \leq 1} \left| \Pi[f_d] - \hat{\Pi}^{\bm{C}_1}_{\text{BQ}}[f_d]\right| 
& \leq & \sqrt{K} \left\|\Pi\left[c_2(\cdot,\bm{x})\right] - \hat{\Pi}^{c_1}_{\text{BQ}}\left[c_2(\cdot,\bm{x})\right]\right\|_{c_2} \; = \;\sqrt{K}  \sup_{\|f\|_{c_2} \leq 1} \left| \Pi[f_d] - \hat{\Pi}^{c_1}_{\text{BQ}}[f_d]\right|.
\end{eqnarray}
We can take $\bm{C}_1$ equal to $\bm{C}_2$ to get: 
\begin{eqnarray*}
\sup_{\|f\|_{\bm{C}} \leq 1} \left| \Pi[f_d] - \hat{\Pi}^{\bm{C}}_{\text{BQ}}[f_d]\right| 
& \leq & \sqrt{K}  \sup_{\|f\|_{c} \leq 1} \left| \Pi[f_d] - \hat{\Pi}^{c}_{\text{BQ}}[f_d]\right|.
\end{eqnarray*}
The convergence for the separable kernel case is therefore driven by the convergence of the scalar-valued kernel. We can therefore use results from the uni-output case in \cite{Briol2015a,Oates2016CF2,Briol2017,Kanagawa2017} to complete the proof. 
\end{proof}

\subsubsection*{Proof of Proposition \ref{cor:sum_separable_kernels}}

\begin{proof}
Note that if the kernel is actually of the form $\bm{C}(\bm{x},\bm{x}') = \sum_{q=1}^Q \bm{B}_q c_q(\bm{x},\bm{x}')$, we can use the triangle inequality satisfied by the norm of $\mathcal{H}_{\bm{C}}$ to show that (for some $C_2>0$):
\begin{eqnarray*}
\sup_{\|f\|_{\bm{C}} \leq 1} \left| \Pi[f_d] - \hat{\Pi}_{\text{BQ}}[f_d]\right| 
& \leq & C_2 \sum_{q=1}^Q \left\|\Pi\left[c_q(\cdot,\bm{x})\right] - \hat{\Pi}_{\text{BQ}}\left[c_q(\cdot,\bm{x})\right]\right\|_{c}^2,
\end{eqnarray*}
so that the overall convergence is dominated by the slowest decaying term.
\end{proof}

\subsection{Proof of Theorem \ref{thm:misspecified_multioutputBQ}}
\begin{proof}
Denote by $\hat{\Pi}^{\bm{C}_\alpha}_{BQ}[\bm{f}]$ the multi-output BQ rule based on $\bm{C}_\alpha$, $\hat{\Pi}^{c_\alpha}_{BQ}[f]$ the uni-output BQ rule based on $c_\alpha$ and $\hat{f}^{\alpha}_d$ the interpolant corresponding this rule. We start by upper bounding the integration error in the uni-output case:
\begin{eqnarray*}
\left|\Pi[f] - \hat{\Pi}^{c_\alpha}_{BQ}[f]\right| 
& \leq & K_1 \|\pi\|_{L_{\infty}(\mathcal{X})}  \|f - \hat{f^{\alpha}}\|_{L_1(\mathcal{X})}  \\
& \leq & K_2 \|f - \hat{f^{\alpha}}\|_{L_2(\mathcal{X})} \\
& \leq & K_3 h^{\beta}_{\bm{X},\mathcal{X}} \rho^{\alpha}_{\bm{X},\mathcal{X}} \|f\|_{L_2(\mathcal{X})}\\
& \leq & K_4 h^{\beta}_{\bm{X},\mathcal{X}} \rho^{\alpha}_{\bm{X},\mathcal{X}} \|f\|_{W_2^{\beta}(\mathcal{X})}\\
& \leq & K_5 h^{\beta}_{\bm{X},\mathcal{X}} \rho^{\alpha}_{\bm{X},\mathcal{X}} \|f\|_{c_{\beta}},
\end{eqnarray*}
for some $K_1,\ldots,K_5 >0$. The first and second inequality correspond to Holder's inequality and the third inequality follows from Theorem 4.2 in \cite{Narcowich2006}. Finally, the fourth and fifth inequalities follow from the definition the Sobolev norm and the norm-equivalence of $\mathcal{H}_{c_\beta}$ and $W_{2}^{\beta}(\mathcal{X})$.

Dividing the above by $\|f_d\|_{\beta}$ on both sides and taking supremums over the unit ball of $\mathcal{H}_{c_\beta}$ we get a result for the worst-case error in the uni-output case:
\begin{eqnarray*}
e(\mathcal{H}_{c_\beta},\hat{\Pi}_{BQ}^{c_\alpha},\bm{X}) \leq K_6 h^{\beta}_{\bm{X},\mathcal{X}} \rho^{\alpha}_{\bm{X},\mathcal{X}}.
\end{eqnarray*}
We can then upper bound the integration error in the multi-output case using Theorem \ref{theorem:convergence_separable} as follows:
\begin{eqnarray*}
\left|\Pi[f_d] - \hat{\Pi}^{\bm{C}_\alpha}_{BQ}[f_d]\right| 
& \leq &  \|\bm{f}\|_{\bm{C}_{\beta}} e(\mathcal{H}_{\bm{C}_\beta},\hat{\Pi}^{\bm{C}_\alpha}_{BQ},\bm{X},d) \\
& \leq & K_6  \|\bm{f}\|_{\bm{C}_{\beta}} e(\mathcal{H}_{c_\beta},\hat{\Pi}_{BQ}^{c_\alpha},\bm{X}) \\
& \leq & K_7   \|\bm{f}\|_{\bm{C}_{\beta}} h^{\beta}_{\bm{X},\mathcal{X}} \rho^{\alpha}_{\bm{X},\mathcal{X}},
\end{eqnarray*}
for some $K_6,K_7>0$.
When $(A_2)$ is satisfied, then we can use the assumption that $h_{\bm{X},\mathcal{X}} \leq C q_{\bm{X}}$ for some constant $C>0$ and the fact that $h_{\bm{X},\mathcal{X}}$ converges as $N^{-\frac{1}{p}}$ to show that the integration error satisfies:
\begin{eqnarray*}
\left|\Pi[f_d] - \hat{\Pi}^{\bm{C}_\alpha}_{BQ}[f_d]\right| 
& \leq &
K_7  \|\bm{f}\|_{\bm{C}_{\beta}} h^{\beta}_{\bm{X},\mathcal{X}} \rho^{\alpha}_{\bm{X},\mathcal{X}}
\; \leq \; K_8   \|\bm{f}\|_{\bm{C}_{\beta}} h^{\beta}_{\bm{X},\mathcal{X}}
\; = \; \mathcal{O}\left(N^{-\frac{\beta}{p}}\right),
\end{eqnarray*}
for some $K_8>0$. 
\end{proof}
\section{Implementation}\label{sec:appendix_implementation}

In this appendix, we present some complementary details which will help users reproduce experiments in the paper.

\subsection{Prior specification}

\subsubsection{Separable kernel} 

The separable matrix-valued kernel is of the form $\bm{C}(\bm{x},\bm{x}') = \bm{B} c(\bm{x},\bm{x}')$ where $c:\mathcal{X}\times \mathcal{X}\rightarrow \mathbb{R}$ is a scalar-valued kernel. If all of the elements $f_d$ of the vector-valued function $\bm{f}$ are evaluated on the same data set $\bm{X}=(\bm{x}_1,\ldots,\bm{x}_N)$, then the Gram matrix can be expressed as 
\begin{eqnarray*}
\bm{C}(\bm{X},\bm{X}) & = & \bm{B} \otimes c(\bm{X},\bm{X}),
\end{eqnarray*}
where $\otimes$ denotes the Kronecker product. Due to properties of the Kronecker, its inverse can then be computed as:
\begin{eqnarray*}
\bm{C}(\bm{X},\bm{X})^{-1}  & = & \bm{B}^{-1} \otimes c(\bm{X},\bm{X})^{-1}.
\end{eqnarray*}
It is straightforward to show that similar expressions can be obtained for the multi-output case of the kernel mean:
\begin{eqnarray*}
\Pi[\bm{C}(\cdot,\bm{X})] & = & \bm{B} \otimes \Pi[c(\cdot,\bm{X})] \; = \; \bm{B} \otimes \left(\int_{\mathcal{X}} c(\bm{x},\bm{X}) \Pi(\mathrm{d}\bm{x}) \right),
\end{eqnarray*}
and initial error:
\begin{eqnarray*}
\Pi\Pi[\bm{C}] 
& = &
\bm{B} \; \Pi\bar{\Pi}[c] \; = \;  \bm{B} \int_{\mathcal{X}\times \mathcal{X}} c(\bm{x},\bm{x}') \Pi(\mathrm{d}\bm{x}) \Pi(\mathrm{d}\bm{x}').
\end{eqnarray*}
These expressions can of course be obtained in closed form whenever the kernel mean and initial error of the scalar-valued kernel are available in closed form. We refer the reader to the table in \cite{Briol2015a} for a list of popular kernels for which this is possible.

\subsubsection{Process Convolution kernel}

In this section, we consider the process convolution kernel given by:
\begin{eqnarray*}
(\bm{C}(x,x'))_{d,d'} & = &  c_{d,d'}(x,x') + c_{w_d}(x,x') \delta_{d,d'}, \\  
c_{d,d'}(x,x')  & = & \sum_{i=1}^{R} \int_{\mathcal{X}} G^i_{d}(x-z) \int_{\mathcal{X}} G^i_{d'}(x'-z') c_i(z,z') \mathrm{d}z' \mathrm{d}z,
\end{eqnarray*} 
This is used in Sec. \ref{sec:application} in the two-output case. There, blurring kernels and reproducing kernels are:
\begin{eqnarray*}
G^1_1(r) & = & \lambda^2_1 \exp\left(-\frac{r^2}{2 \sigma_1^2}\right), \\
G^1_2(r) & = & \lambda^2_2 \exp\left(-\frac{r^2}{2 \sigma_2^2}\right), \\
c_1(x,y) & = & \lambda^2_3 \exp\left(-\frac{(x-y)^2}{2 \sigma_3^2}\right), \\
G^2_1(r) & = & \lambda^2_4 \exp\left(-\frac{r^2}{2 \sigma_4^2}\right), \\
G^2_2(r) & = &  \lambda^2_5 \exp\left(-\frac{r^2}{2 \sigma_5^2}\right), \\
c_{2}(x,y) & = & \lambda^2_6 \exp\left(-\frac{(x-y)^2}{2 \sigma_6^2}\right),
\end{eqnarray*}
for some constants $\sigma_i,\lambda_i>0$ for $i=1,\ldots,6$. Note that for simplicity, we did not include $c_{w_1}$ and $c_{w_2}$. The kernel mean and initial error can easily be computed in closed form using Gaussian identities.

\subsection{Hyper-parameters}

One of the main challenges when using uni-output BQ and multi-output BQ is the selection of appropriate hyperparameters. In this section, we consider multi-output BQ with $\mathcal{GP}(\bm{0},\bm{C})$ prior we denote the hyperparameters of the kernel $\bm{C}$ in vector form as $\bm{\theta}=(\theta_1,\ldots,\theta_l)$.
To optimise these parameters, we propose to use an empirical-Bayes approach and maximise the log-marginal likelihood:
\begin{eqnarray*}
\log p \left(\bm{f}(\bm{X})|\bm{X},\bm{\theta} \right) & = & -\frac{1}{2}\bm{f}(\bm{X})^\top \bm{C}(\bm{X},\bm{X})^{-1}\bm{f}(\bm{X}) - \frac{1}{2}\log \left|\bm{C}(\bm{X},\bm{X})\right|  -\frac{ND}{2}\log (2 \pi).
\end{eqnarray*}
This can be efficiently optimised by making use of gradients of the log-marginal likelihood $\forall i \in \{1,\ldots,l\}$:
\begin{eqnarray*}
\frac{\partial \log p \left(\bm{f}(\bm{X})|\bm{X},\bm{\theta} \right)}{\partial \theta_i} 
& = &
\frac{1}{2} \bm{f}(\bm{X})^\top \bm{C}(\bm{X},\bm{X})^{-1} \frac{\partial \bm{C}(\bm{X},\bm{X})}{\partial \theta_i} \bm{C}(\bm{X},\bm{X})^{-1} \bm{f}(\bm{X}) - \frac{1}{2} \text{Tr}\left(\bm{C}(\bm{X},\bm{X})^{-1}\frac{\partial \bm{C}(\bm{X},\bm{X})}{\partial \theta_i}\right).
\end{eqnarray*}

\section{Extended Numerical Experiments}\label{sec:appendix_numerical_exp}

In this appendix, we provide additional results for the multi-output BQ experiments provided in Sec. \ref{sec:application} for the multifidelity toy models and the global illumination problem. We also include numerical results for a popular variational approximation of multi-output GP.

\subsection{Scaling multi-output BQ with variational approximations}

Computational burdens are heavy for multi-output BQ due to the inversion of $ND\times ND$ matrix. The computational complexity is $O(N^3D^3)$ and the storage is $O(N^2D^2)$. \cite{Alvarez2011convolved} introduced and fully discussed a sparse approximation of multi-output GPs with process convolution kernels, using the fact that outputs are conditionally independent if the latent functions is fully observed. This idea can then be extended to multi-output BQ by taking our posterior on the value of the integrals as the pushforward through the integral operator of the approximate multi-output GP.

\begin{figure}[htb!]
\centering
\includegraphics[scale=0.18]{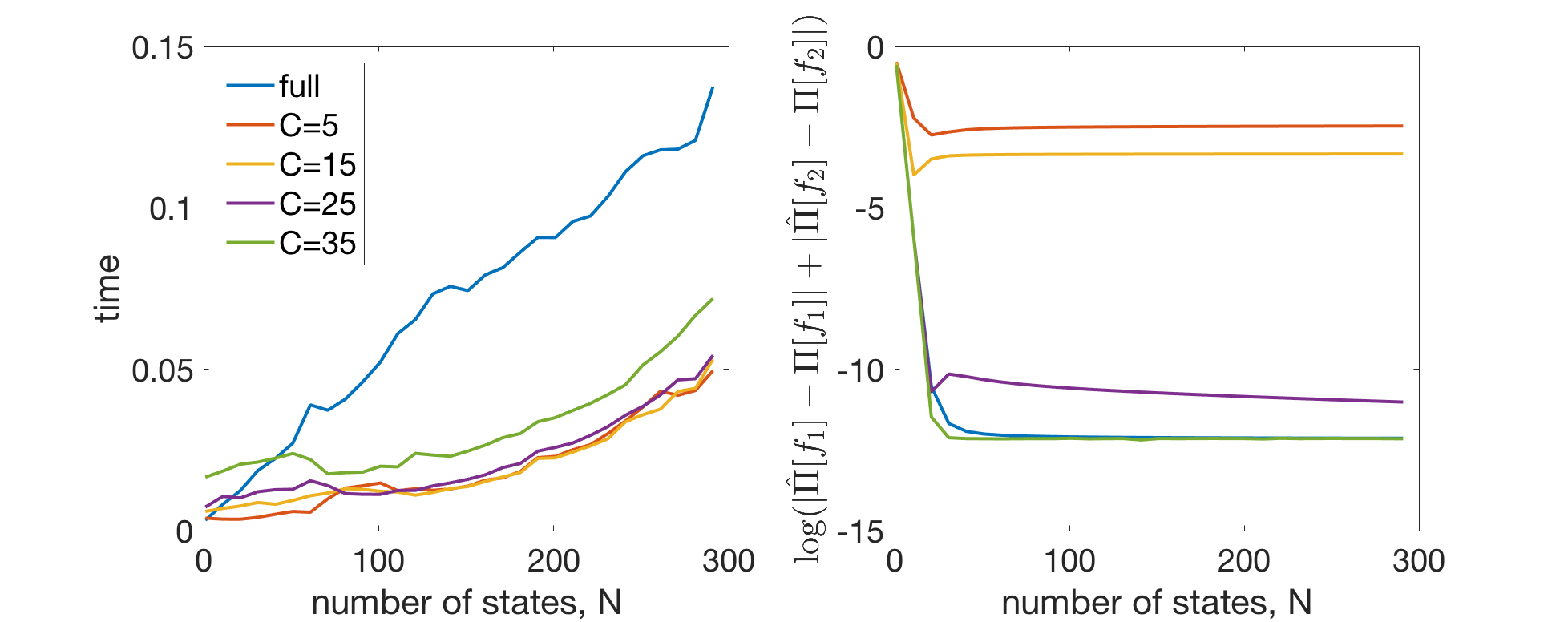} 
\caption{Variational approximation: Plot of the computational times (in seconds) and log-errors of full/ approximated multi-output BQ against the number of points given for different number of points evaluated from the latent function.}
\label{fig: figure_5_1_appendix}
\end{figure}

 Consider functions $f_1(x)=3\cos(\frac{\pi x}{5})$ and $f_2(x)=0.7\cos(\frac{2.2\pi x}{10})$ on $\mathcal{X}=[-5,5]$. Computational times and log-errors of multi-output BQ estimates for integrals of these functions against a uniform measure $\Pi$ with and without the variational approximation by \cite{Alvarez2011convolved} using different number of equidistant points between $-5$ and $5$ are given in Fig. \ref{fig: figure_5_1_appendix}. This approximation is considered for different number of points evaluated from the latent function, i.e. $C=5, 15, 25, 35$. Regarding the process convolution kernel, $G_1^1$, $G_1^2$, $c_1$, $G_2^1$, $G_2^2$, $c_2$ are squared-exponential kernels with amplitude and lengthscale parameters ($\sqrt{3}, 1.3$), ($0.7, 1$), $(1,1)$, ($0.9, 0.6$), ($0.6, 0.5$) and ($0.8, 1$) respectively.

Clearly, with a large enough number of points $C$, the same integration accuracy as for the full GP can be obtained at much lower computation cost. This could make variational approximations a promising approach for multi-output BQ, but this would warrant a much more extensive study.

\subsection{Multi-fidelity modelling}\label{sec:appendix_multifidelity}

\begin{figure}[h!]
\begin{center}
\includegraphics[trim={0 0 0 0},clip,width=0.8\textwidth]{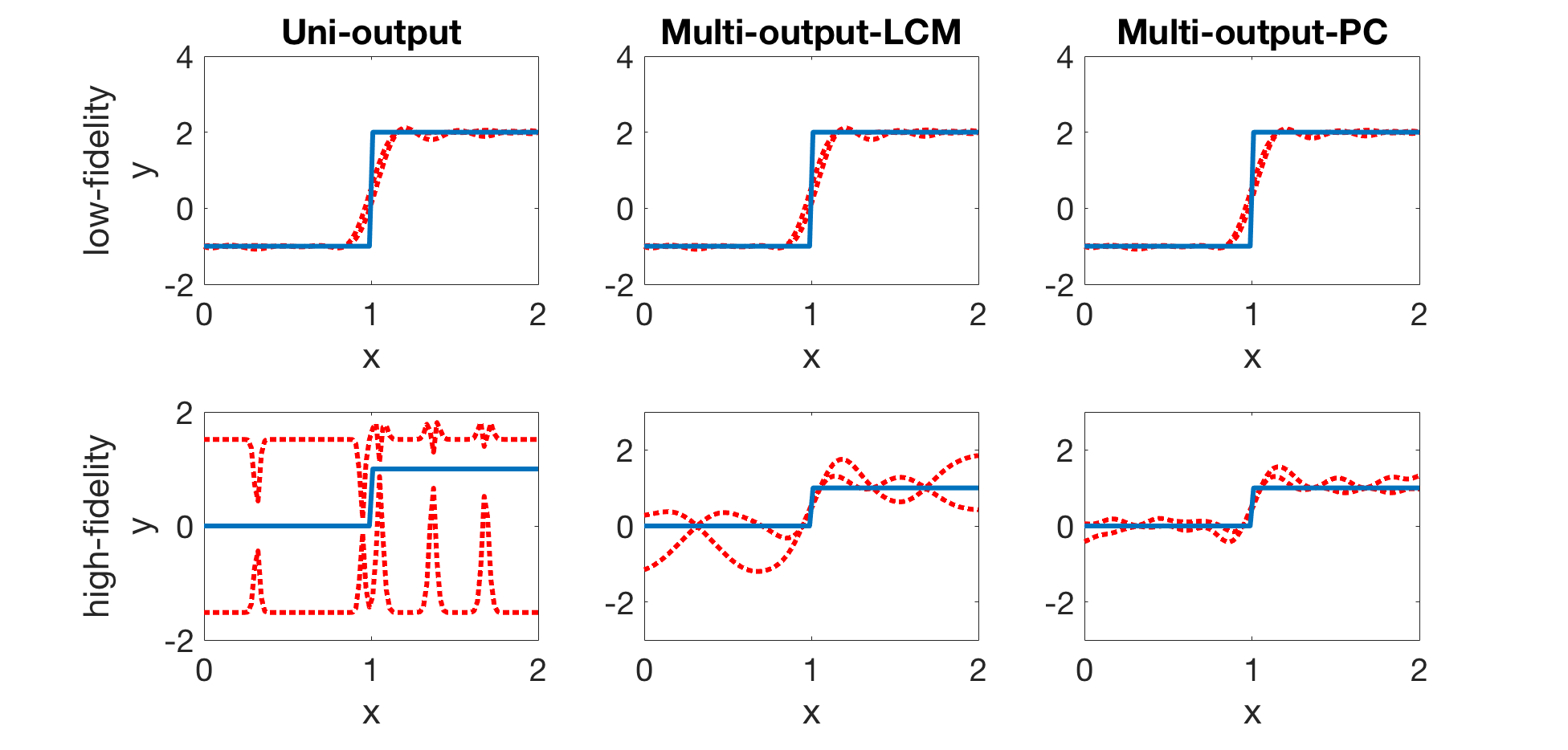}\\
\vspace{2mm}
\includegraphics[trim={0 0 0 0},clip,width=0.8\textwidth]{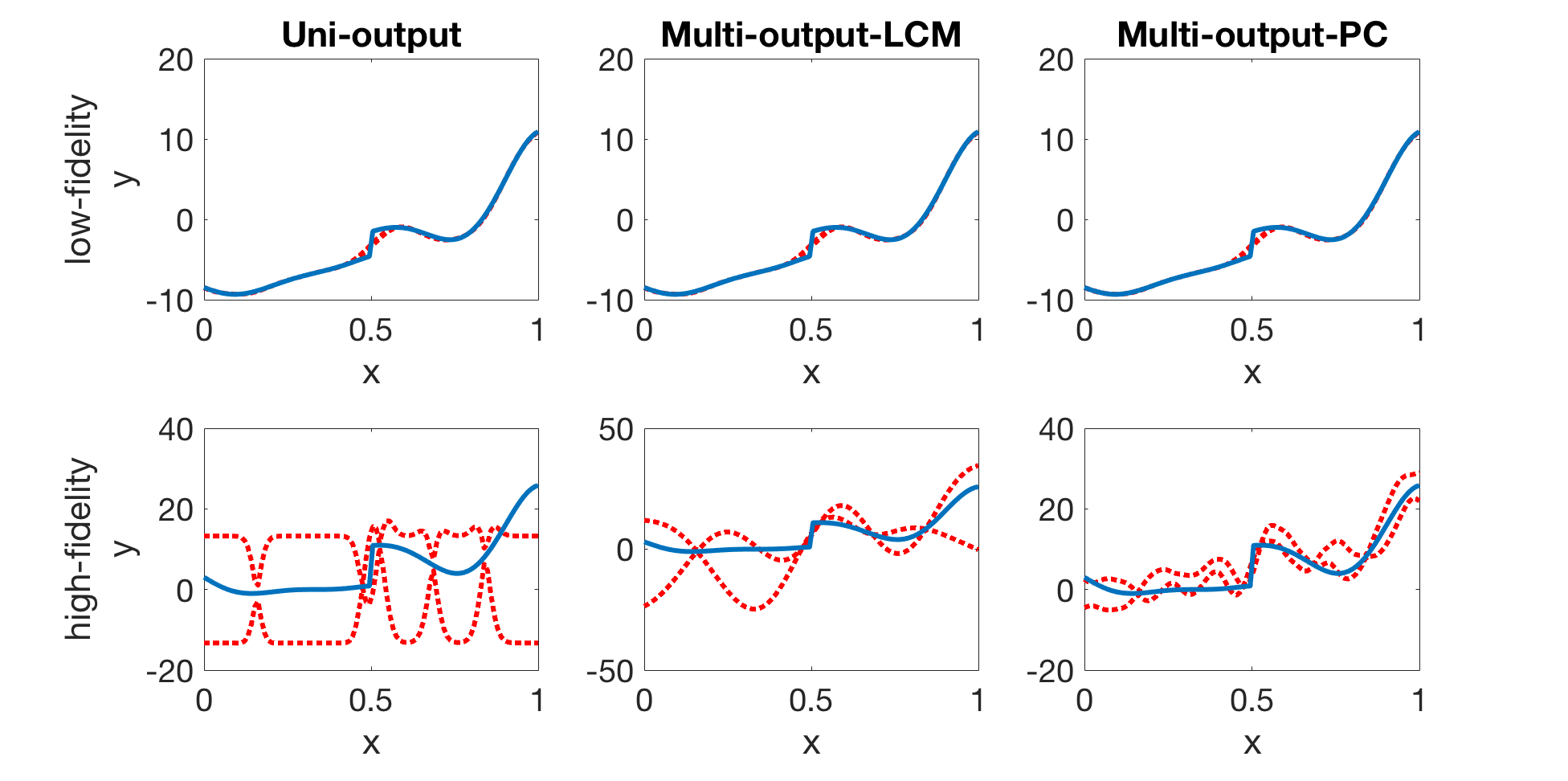}
\caption{Multi-fidelity modelling: Plot of the Step function (top) and Forrester function (bottom) in blue.  The Gaussian Process 95\% credible intervals are given by dotted red lines in each case.}
\label{fig:mulitifidelity_credibleints}
\end{center}
\end{figure}

In Fig. \ref{fig:mulitifidelity_credibleints}, we give an extended version of Fig. \ref{plot:multifidelity_functions} which includes credible intervals for each of the multi-output GP models. For both functions, the high-fidelity confidence intervals in the uni-output case are overly pessimistic, whereas for the multi-output cases, the posterior is concentrated on the true functions. One interesting point is that both of the multi-output BQ methods are over-confident near the kinks in the functions. This is to be expected since the true functions do not like in the RKHS corresponding to the kernel used for these BQ rules.

As an extension to these experiments, we consider a steady-state version of the Allen-Cahn equation on $[0,10]$ subject to a sinusoidal forcing term and with boundary conditions:
\begin{equation} \label{eq:allencahn1}
\epsilon\frac{\partial^2 u}{\partial x^2}+u-u^3 \; = \; \sin(x),\ \ \  u(0)\; = \; 1, \quad u(10) \; = \; -1,
\end{equation} 
where $\epsilon$ controls the rate of diffusion \cite{allencahn}. Our target is to approximate the integral of the solution of Eq. \ref{eq:allencahn1} for $\epsilon\approx 0$. We take solutions of Eq. \ref{eq:allencahn1} on $\mathcal{X}=[0,10]$ with $\epsilon=2$ and $\epsilon=0.1$ as our low-fidelity model and high-fidelity model respectively. Ideally, we would prefer to take $\epsilon$ as small as possible but this complicates the numerical approximation of the solution.

\begin{figure}[htb!]
\centering
\includegraphics[scale=0.2]{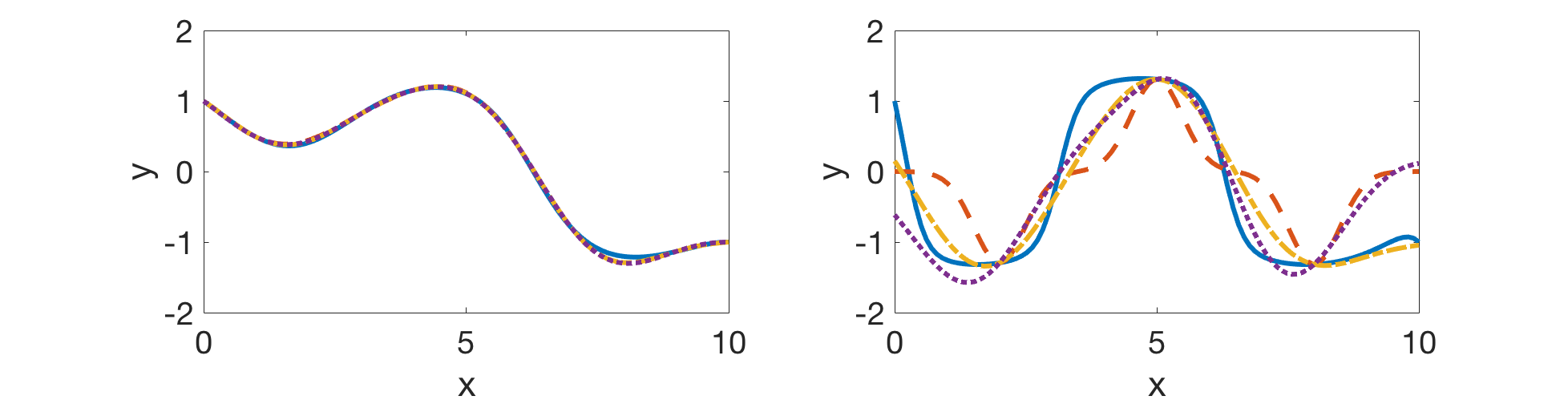} 
\caption{Multi-fidelity modelling: Plot of the solutions of Equation \ref{eq:allencahn1} for $\epsilon=2$ (left) and $\epsilon=0.1$ (right). Each plot gives the true function (blue) and their uni-output (dashed, red), LMC-based multi-output (dashed, yellow) and PC-based multi-output (dotted purple) approximations.}
\label{fig: figure_1_3_appendix}
\end{figure}

\begin{figure}[htb!]
\centering
\includegraphics[scale=0.18]{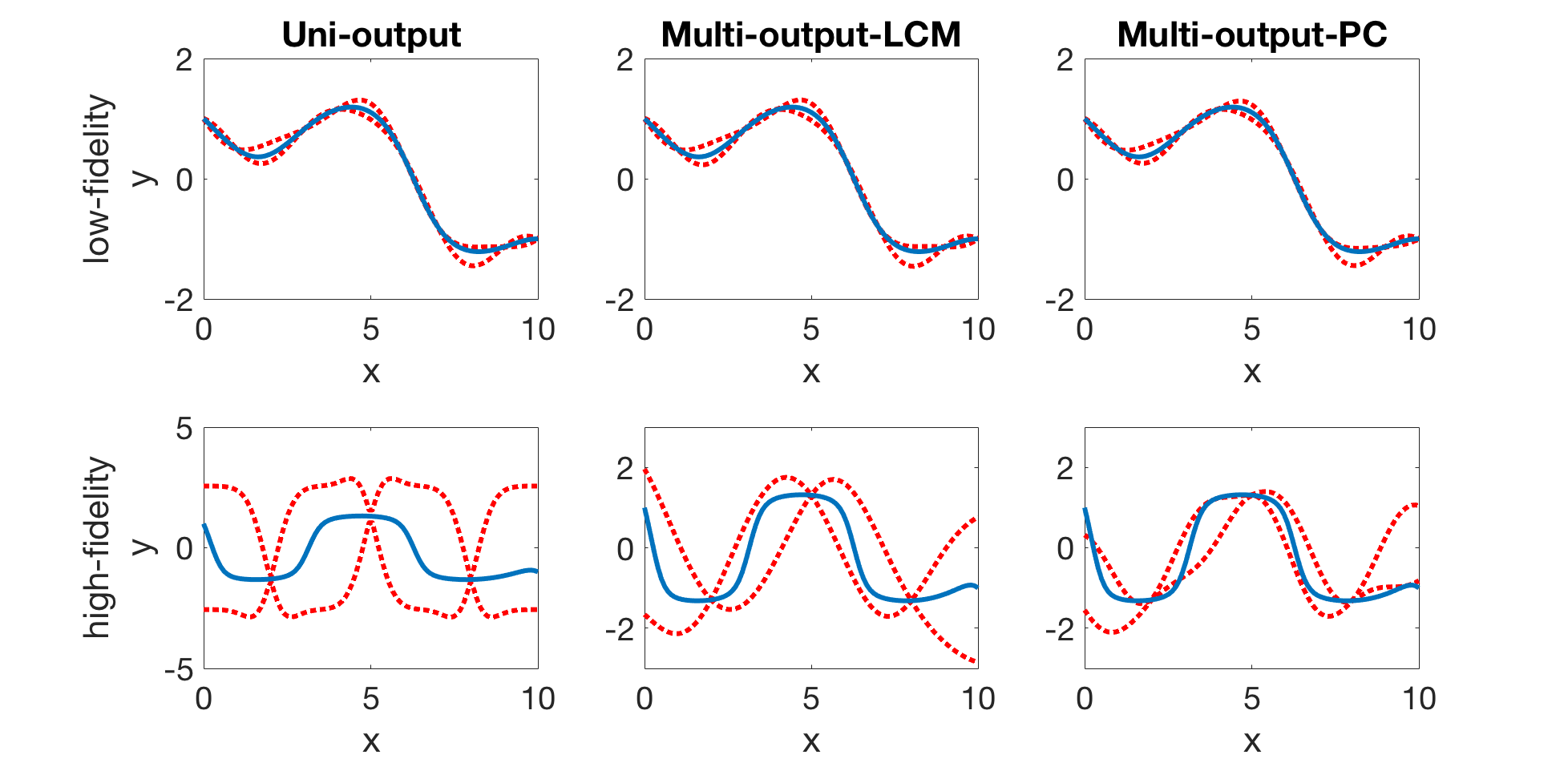} 
\caption{ Multi-fidelity modelling: Plot of the solutions of Equation \ref{eq:allencahn1} for $\epsilon=2$ (top) and $\epsilon=0.1$ (bottom) in blue. The Gaussian Process 95\% credible intervals are given by dotted red lines in each case.}
\label{fig: figure_6_3_appendix}
\end{figure}

 The functions considered and corresponding posteriors are given in Fig. \ref{fig: figure_1_3_appendix} and \ref{fig: figure_6_3_appendix}, while the uni-output and multi-output BQ estimates for integration of these functions against a uniform measure $\Pi$ are given in the table in Fig. \ref{table:multifidelity_appendix}. Integer points between 0 and 10 are evaluated, with points at 2, 5 and 8 being used to evaluate the high fidelity model and the others used for the low fidelity model. The choice of kernel hyperparameters is made by maximising the marginal likelihood. Clearly, the two multi-output BQ algorithm give posteriors on the high-fidelity model which are much more concentrated on the true function than the uni-output BQ.

\begin{figure}[htb!]
\begin{center}
\begin{tabular}{|c|c|c|c|} 
 \hline
 Model & BQ & LMC-BQ & PC-BQ \\  [0.5ex] 
 \hline
 AC (l) & 0.004 (0.197) & 0.006 (0.187) & 0.007 (0.388)\\ 
 AC (h) & 0.211 (0.27) & 0.002 (0.444) & 0.037 (0.191)\\ 
 \hline
\end{tabular}
\end{center}
\caption{Multi-fidelity modelling: Performance of uni-output BQ and multi-output BQ (with LMC and PC kernels) on the Allen-Cahn problem (AC) both for the low fidelity (l) and high fidelity (h) cases.}
\label{table:multifidelity_appendix}
\end{figure}


\subsection{Global illumination problem}\label{sec:appendix_global_illum}

\begin{figure}[h!]
\begin{center}
\includegraphics[trim={2cm 0 2cm 0},clip,width=0.52\textwidth]{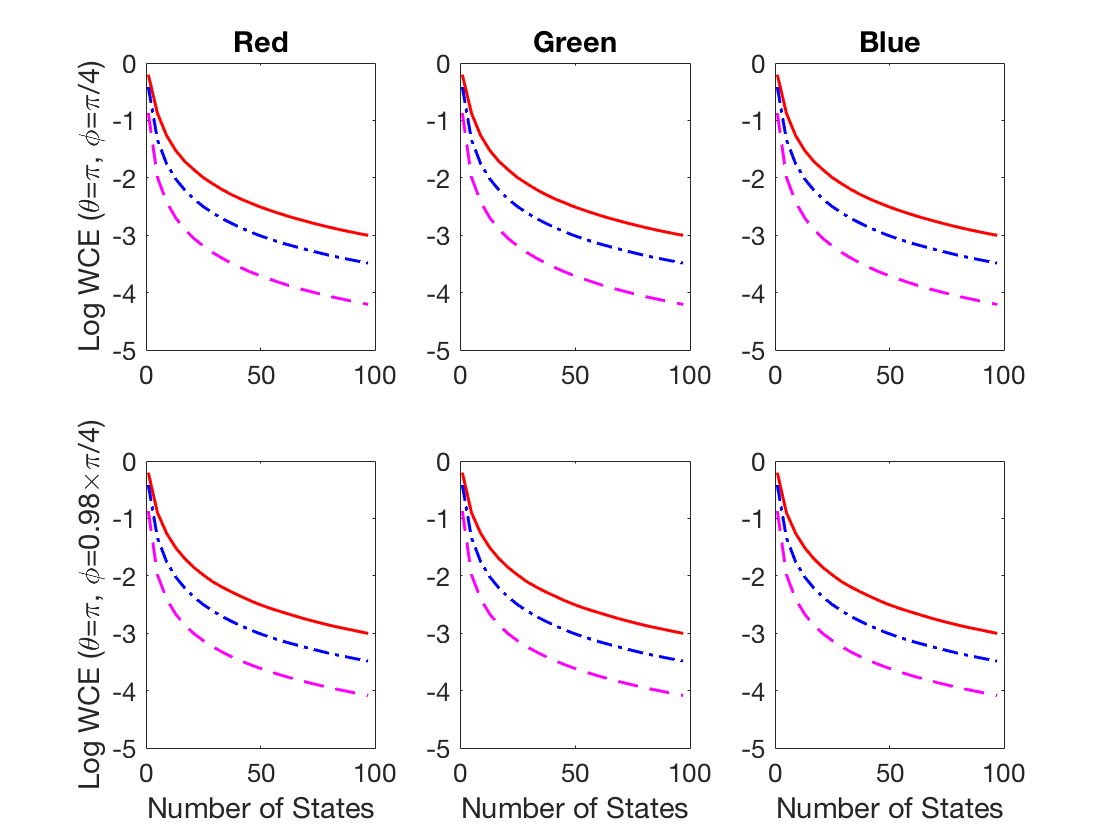}
\caption{Global illumination: Plot of the worst-case integration error for $f_1,f_2$ in the case of the red, blue and green channels. Uni-output BQ is given in red (full line) whilst two-output BQ based on LMC is given in blue (dashed and dotted line) and five-output BQ based on LMC is given in magenta (dashed line).}
\label{plot:global_illum_WCE}
\end{center}
\end{figure}

In Fig. \ref{plot:global_illum_WCE}, we plot the evolution of the worst-case integration error as $N$ increases for the uni-output, two-output and five-output BQ with LMC kernel. As expected from Thm \ref{theorem:convergence_separable}, the convergence occurs at the same rate in $N$ but with a smaller rate constant the more outputs there are.

\end{document}